\documentclass[pre]{revtex4}
\usepackage[dvips]{graphicx}
\usepackage{amssymb,amsthm}
\usepackage{multirow}

\newcommand{\beq}[0]{\begin{equation}}
\newcommand{\eeq}[0]{\end{equation}}
\newcommand{\e}{\varepsilon}

\newcommand{\la}{\langle}
\newcommand{\ra}{\rangle}
\newcommand{\ds}{\displaystyle}
\newcommand{\avh}{\la H_1 \ra}
\newcommand{\pa}{\partial}
\newcommand{\w}{\omega}
\newcommand{\ud}{{\mathrm d}}
\newcommand{\bc}{\begin{center}}
\newcommand{\ec}{\end{center}}
\newcommand{\lesq}{\leqslant}
\newcommand{\gesq}{\geqslant}
\newcommand{\xz}{{\chi_z}}

\begin{document}
\author{V. Koukouloyannis}
\affiliation{Department of Civil Engineering, Technological Educational Institute of Serres, 62124 Serres, Greece}
\author{P.G. Kevrekidis}
\affiliation{Department of Mathematics and Statistics, University of
Massachusetts, Amherst MA 01003-4515}
\author{J. Cuevas}
\affiliation {Nonlinear Physics Group. Departamento de F\'isica Aplicada I, Escuela Polit\'ecnica Superior. Universidad de Sevilla. C/ Virgen de \'Africa, 7, 41011 Sevilla, Spain}
\author{V. Rothos}
\affiliation {Department of Mathematics, Faculty of Engineering, Aristotle University of Thessaloniki,Thessaloniki GR54124 Greece}

\title{Multibreathers in Klein-Gordon chains with interactions beyond
nearest neighbors}

\theoremstyle{plain} \newtheorem{theorem}{Theorem}

\begin{abstract}
We study the existence and stability of multibreathers in Klein-Gordon chains
with interactions that are not restricted to nearest neighbors. 
We provide a general framework where
such long range effects can be taken into consideration for arbitrarily
varying (as a function of the node distance) linear couplings between
arbitrary sets of neighbors in the chain. By examining special case
examples such as three-site breathers with
next-nearest-neighbors, we find {\it crucial} modifications
to the nearest-neighbor picture of one-dimensional oscillators being excited
either in- or anti-phase. 
Configurations with nontrivial phase profiles emerge from or collide with the ones with standard
($0$ or $\pi$) phase difference profiles, through supercritical or subcritical
bifurcations respectively. Similar bifurcations emerge when examining four-site breathers
with either next-nearest-neighbor or even interactions with the three-nearest
one-dimensional neighbors. The latter setting can be thought of as a prototype
for the two-dimensional building block, namely a square of lattice nodes,
which is also examined. Our analytical predictions are found to be in
very good agreement with numerical results.
\end{abstract}

\maketitle

\section{Introduction}

The initial numerical inception of anharmonic modes consisting
of a few excited sites in nonlinear lattices~\cite{sietak,pa90},
and their subsequent placement on a rigorous existence basis (under
rather generically satisfied non-resonance conditions) in~\cite{macaub}
has triggered a huge growth of interest in the theme of the
so-called discrete breathers. These are exponentially localized
in space, periodic in time states which have subsequently been
theoretically/numerically predicted and experimentally verified
to arise in a very diverse host of applications. These include
(but are not limited to) DNA double-strand dynamics in
biophysics \cite{peyrard},
coupled waveguide arrays and photorefractive crystals
in nonlinear optics \cite{photon,moti3,review_opt}, breathing oscillations
in micromechanical cantilever arrays \cite{sievers},
Bose-Einstein condensates in optical lattices in atomic
physics \cite{morsch}, and
granular crystals \cite{sen08}. The interest in this theme has also
been mirrored in a wide array of reviews on methods of identifying and
analyzing such intrinsically localized
modes~\cite{flach1,macrev,flach2,aubrev}.

More recently, the stability of the discrete breather configurations,
especially in the case of the excitation of multiple sites has been
of particular interest. One approach to obtaining relevant results
consists of the so-called Aubry band theory~\cite{aubrev},
used e.g. in~\cite{ACSA03,CAR05}. This led to the conclusion that for soft nonlinear
potentials multi-breathers with any subset of adjacent sites being
excited in-phase are unstable, while ones with all
adjacent sites in anti-phase can be stable in the vicinity of the
so-called anti-continuum limit of uncoupled anharmonic oscillators.
A complementary theory that yields insights on both the existence
and the stability of multibreathers has been pioneered by MacKay
and collaborators; see e.g., \cite{mackay1,mac1,macsep}. This is the
so-called effective Hamiltonian method which is identified by averaging over
the period of the unperturbed solution and developing the proper action-angle
variables. The extrema of the resulting effective Hamiltonian determine
the relative phases of adjacent excited sites in the multi-site breather
solution, while the relevant Hessian is intimately connected to the
Floquet multipliers of the associated periodic orbit. Using this methodology,
the work of~\cite{KK09} retrieved as well as refined the results of~\cite{ACSA03}
for arbitrary phase relations between the excited oscillators of such
multi-breather configurations. The equivalence between these two
basic methods and their conclusions was recently established in~\cite{IJBC}.
We should also note in passing that similar results have been acquired also in configurations where there are ``holes'' between the excited breather sites~\cite{pelisak}, through higher order perturbation theory
generalizing the above conclusions to the cases with one-site holes. On the other hand, the existence and stability of single/multi-site breathers have been studied in diatomic FPU lattices. The work of~\cite{yoshi} was based
on a discrete Sturm theorem which necessitated (for the separation of
the space $n$ and time $t$ variables) a potential which was at least
purely quartic. In the realm of lattices with longer than the 
nearest-neighbor interactions a variety of issues have been considered such as, 
e.g. in \cite{FeckanRothos10}, the existence and bifurcation of quasi 
periodic traveling waves in nonlocal lattices with polynomial type 
potentials.

In the present work, we consider the generalization of the above
settings, which are principally concerned with the interaction between
nearest neighbors, to the case with longer range neighbor interactions
for Klein-Gordon chains.
Upon revisiting the nearest neighbor case and presenting the
effective Hamiltonian formalism (of MacKay and collaborators) there
(section II) for existence and stability of multibreathers, in section
III, we generalize this formalism to the case of an arbitrary number
of neighbors (denoted by $r$) interacting with each other.
By specializing to the case of nearest and next-nearest neighbor
interactions (and three-site breathers)
as our first case example of the application of the
results in section IV, we already infer the fundamental modifications
to the standard picture that ensue due to interactions beyond nearest
neighbors.
These include configurations that have {\it non-standard relative
phases between adjacent oscillators}, a feature which is absent
in the nearest-neighbor interaction case \cite{kouk12} and also {\it symmetry breaking
bifurcations} that arise due to the ``collision'' of branches
of solutions with such non-trivial phase relations, with more standard
ones with relative phases of $0$ or $\pi$ between adjacent oscillators.
The generic nature of these conclusions is confirmed by considering
the case examples of four-site breathers with next-nearest-neighbor
interactions in section V and such breathers with interaction ranges
of $r=3$ in section VI. The latter setting is very close to genuinely
two-dimensional setting in a square lattice plaquette, which constitutes
our final example in section VII. We close our presentation by some
remarks on the parallels of our results with the simpler case of the
discrete nonlinear Schr{\"o}dinger lattices~\cite{pgkbook} (section VIII),
which has been examined earlier in~\cite{pgkpla,chong}, as well as a summary of
our conclusions and some future directions (section IX).


\section{Background: The Classical Klein-Gordon chain}

\begin{figure}[htbp]
	\centering
		\includegraphics[width=10cm]{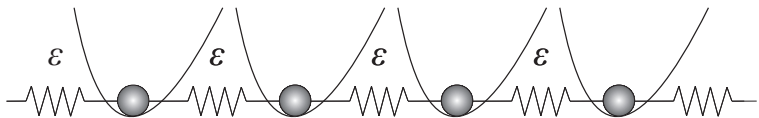}
		\caption{The classical nearest-neighbor Klein-Gordon chain}
	\label{fig:KGchain}
\end{figure}

The Hamiltonian of a Klein-Gordon chain with nearest neighbor interactions is the following
\beq H=H_0+\e H_1=\sum_{i=-\infty}^{\infty}\left[ \frac{1}{2}p_i^2 +V(x_i)\right] +\frac{\e}{2}\sum_{i=-\infty}^{\infty}\left(x_i-x_{i-1}\right)^2,\label{ckg}\eeq
which leads to the equations of motion
$$\ddot{x_i}=-V'(x_i)+\e(x_{i-1}-2x_i+x_{i+1}).$$

It is well known that this system supports discrete breather, as well as,
multibreather solutions. As indicated above, there are several papers
dealing with the existence and stability of these motions; see e.g. \cite{macaub, sepmac, koukicht1, koukmac,mackay1, ACSA03, KK09, pelisak}.

\subsection{Persistence of mutibreathers}
In the anti-continuum limit $\e=0$, we consider all the oscillators of the chain at rest except for $n+1$ ``central'' ones which move in periodic orbits of
frequency $\w$. As indicated in~\cite{pelisak}, it is possible to generalize
considerations to the case where not all of these oscillators are adjacent
to each other, however, we will not concern ourselves with this additional
complication herein.
The time-periodic and space-localized motion of our excited oscillators
will persist for $\e\neq0$ to provide multibreathers of the same frequency $\w$, if the phase difference between the succesive central oscillators satisfies
specific conditions. In \cite{mackay1} it was shown that multibreathers
correspond to critical points of $H^{\mathrm{eff}}$ which in first order of
approximation is given by $H^{\mathrm{eff}}=H_0(I_i)+ \e \avh(\phi_i, I_i)$ \cite{koukmac}. The variables $\phi_i=w_{i+1}-w_i$ denote the $n$ phase differences of the $n+1$ successive central oscillators, while $I_i$ are given by $I_i=\sum_{j=i}^{n} J_j$, where $(J_i, w_i)$ are the action-angle variables of each uncoupled oscillator.

The average value of the coupling part of the Hamiltonian
$$\avh(\phi_i, I_i)=\frac{1}{T}\oint H_1(w_0, \phi_i, I_i)\ud t$$
is calculated along the orbits in the anti-continuum limit $\e=0$.

This yields the conclusion that the persistence conditions for the existence of $n+1$-site multibreathers are
\beq\frac{\pa \avh}{\pa \phi_i}=0,\quad i=1\ldots n,\label{gen_per}\eeq
Note that the persistence conditions are the same for every lattice case where the Hamiltonian can be written in the for $H=H_0+\e H_1$ with $\frac{\pa \avh}{\pa \phi_i} \not\equiv 0$. For a more detailed description of the above procedure one can also see \cite{KK09}.

The motion of the central oscillators for $\e=0$ can be described by
\beq x_i(w_i)=\sum_{m=0}^\infty A_m(J_i)\cos(mw_i).\label{xdevel}\eeq
Since  the action $J_i$ remains constant along an orbit in the anticontinuum limit, $x_i$ depends only on $w_i$. So, 
the average value of $H_1$ becomes (\cite{KK09} appendix A)
$$\avh=-\frac{1}{2}\sum_{m=1}^\infty\sum_{s=1}^{n}A_m^2\cos(m\phi_s)$$
and the persistence conditions (\ref{gen_per}) become in the case of Klein-Gordon chains with nearest neighbor interactions,
\beq\frac{\pa \avh}{\pa \phi_i}=0\Rightarrow M(\phi)\equiv\sum_{m=1}^{\infty}mA_m^2\sin(m\phi_i)=0,\quad i=1\ldots n.\label{per_r1}\eeq
The function $M(\phi)$ possesses the obvious solutions $\phi_i=0, \pi$, while
it has no others, as it is shown in \cite{kouk12}.

\subsection{Stability of multibreathers}
The spectral stability of the above mentioned multibreather solutions or,
equivalently, the linear stability of the corresponding periodic orbits is determined through its {\it characteristics exponents} $\sigma_i$. These exponents are connected with the corresponding Floquet multipliers by the relation
$$\lambda_i=e^{\sigma_i T},$$
where  $T=2\pi/\w$ is the period of the multibreather. Due to the Hamiltonian
character of the system there is a pair of exponents identically
equal to zero. The non-zero characteristic exponents of the central
oscillators correspond to the
eigenvalues of the $(2n\times 2n)$ {\it stability matrix}
\cite{mackay1}
${\bf E}={\bf \Omega}
D^2H^{\mathrm{eff}}$, where $\Omega$ is the matrix of the symplectic form
${\bf \Omega}=\left(\begin{array}{cc}\bf O&-\bf I\\\bf I&\bf O\end{array}\right)$ and $\bf I$ is the
$n\times n$ identity matrix. The effective Hamiltonian $H^{\mathrm{eff}}$, as it has already been mentioned, in first order of approximation is given by $H^{\mathrm{eff}}=H_0+\e\avh$. So, the stability matrix $\bf E$, to leading order of approximation and by
taking into consideration the form of $H^{\mathrm{eff}}$, becomes
\beq \bf E=\left(\begin{array}{c|c}\bf A&\bf B\\ \hline\bf C&\bf D\end{array}\right)=\left(\begin{array}{c|c}\e\bf A_1&\e\bf B_1\\ \hline\bf C_0+\e\bf C_1&\e\bf D_1\end{array}\right)=\left(\begin{array}{c|c}
-\e\ds\frac{\pa^2\avh}{\pa\phi_i\pa I_j}&-\e\ds\frac{\pa^2\avh}{\pa\phi_i\pa\phi_j}\\[10pt]
\hline\\[-8pt]
\ds\frac{\pa^2H_0}{\pa I_i I_j}+\ds\e\frac{\pa^2\avh}{\pa I_i\pa I_j}&\ds\e\frac{\pa^2\avh}{\pa\phi_j\pa I_i}
\end{array}\right).\label{e1}\eeq

Since the only possible solutions are the ones with $\phi_i=0, \pi$ and we consider central sites oscillating with the same frequency $\w$, we get that $\bf A_1=D_1=0$ and so, the nonzero characteristic exponents are given to leading order of approximation by
\beq\sigma_{\pm i}=\pm\sqrt{\e\,\chi_{1i}}+{\cal O}(\e^{3/2})\quad i=1\ldots n\label{sx},\eeq
where $\chi_{1i}$ are the eigenvalues of the matrix $\bf B_1\cdot C_0$.
Due to the form of the $J\mapsto I$ transformation the $\bf C_0$ matrix becomes (see \cite{KK09} appendix B)
$${\bf C_0}=\frac{\pa^2 H_0}{\pa I_i\pa I_j}=-\frac{\pa\w}{\pa J}\cdot{\bf L}=-\frac{\pa\w}{\pa J}\cdot
\left(\begin{array}{ccccc}
2&-1&0 & & \\
-1&2&-1&0 & \\
 &\ddots&\ddots&\ddots & \\
 & 0 &-1&2&-1\\
 &  &0 & -1&2
\end{array}\right).$$
So, (\ref{sx}) becomes, up to leading order terms,
\beq\sigma_{\pm i}=\pm\sqrt{\ds-\e\frac{\pa \w}{\pa J}{\chi_z}_i}\quad i=1\ldots n\label{sz},\eeq
where ${\chi_z}_i$ are the eigenvalues of ${\bf Z}={\bf B_1\cdot L}$.

For systems of the form (\ref{ckg}) we get
$${\bf B_1}=\ds\frac{\pa^2\avh}{\pa\phi_i\pa\phi_j}=
\left\{\begin{array}{cl}
f_i&\mathrm{for}\ i=j\\
0& \mathrm{for}\ i\neq j
\end{array}\right. ,$$
with
\beq f_i=f(\phi_i)=\frac{1}{2}\sum_{m=1}^{\infty}m^2A_m^2\cos(m\phi_i)\label{eq:f}.\eeq
So, $\bf Z$ can be written as
\beq {\bf Z}={\bf B_1}\cdot{\bf L}=
\left(\begin{array}{ccccc}
f_1&0&0 & & \\
0&f_2&0&0 & \\
 &\ddots&\ddots&\ddots & \\
 & 0 &0&f_{n-1}&0\\
 &  &0 & 0&f_n
\end{array}\right)\cdot
\left(\begin{array}{ccccc}
2&-1&0 & & \\
-1&2&-1&0 & \\
 &\ddots&\ddots&\ddots & \\
 & 0 &-1&2&-1\\
 &  &0 & -1&2
\end{array}\right)
=\left(\begin{array}{ccccc}
2f_1&-f_1&0 & & \\
-f_2&2f_2&-f_2&0 & \\
 &\ddots&\ddots&\ddots & \\
 & 0 &-f_{n-1}&2f_{n-1}&-f_{n-1}\\
 &  &0 & -f_n&2f_n
\end{array}\right)\label{zz}\eeq

Note that, for linear stability we require all the Floquet multipliers to lie on the unit circle, which is tantamount to all the characteristic exponents
being purely imaginary. This depends on the sign of $P=\e\frac{\pa \w}{\pa J}$ and the sign of $\chi_z$ as it can be seen from (\ref{sz}). Finally, by using some counting theorems \cite{KK09} for (\ref{zz}), we obtain:

\begin{theorem}\label{thm1}~\cite{KK09}
In systems of the form (\ref{ckg}), if $P\equiv\e\frac{\pa \w}{\pa J}<0$ the only configuration which leads to
linearly stable multibreathers, for $|\e|$ small enough, is the one
with $\phi_i=\pi\quad\forall i=1\ldots n$ (anti-phase
multibreather), while if $P>0$ the only linearly
stable configuration, for $|\e|$ small enough, is the one with
$\phi_i=0\quad\forall i=1\ldots n$ (in-phase multibreather).
Moreover, for $P<0$ (respectively, $P>0$), for unstable configurations, their number of
unstable eigenvalues will be precisely equal to the number of nearest
neighbors which are in- (respectively, in anti-) phase between them.
\end{theorem}

\noindent {\bf Remark 1:} Note that, the form of the matrix $\bf C_0$ is such
due to the form of the $J\mapsto I$ transformation and the fact that
in the anti-continuum limit $\frac{\pa H_0}{\pa J}=\w$ and $\w_i=\w\ \mathrm{for}\ i=1\ldots n$. So, it is independent of the range of the interaction between the oscillators of the chain and it will remain the same in what follows. On the other hand, the diagonal form of $\bf B_1$ will change if longer range interactions are added to the system. So, the theorem will {\it no longer hold} but the general methodology will still apply and the characteristic exponents of the multibreather will be given by (\ref{sz}). We will consider
this case in what follows.

\noindent {\bf Remark 2:} In our previous works we used the term ``out-of-phase'' for $\phi=\pi$ configurations. This was because the only out-of-phase configuration was the $\phi=\pi$ one. In the present work, since, as we will see in the next section, there are out-of phase configurations with $\phi\neq\pi$, we use the term ``anti-phase'' for the $\phi=\pi$ configuration.

\section{Klein-Gordon chain with long range interactions}

\begin{figure}
	\centering
		\includegraphics[width=10cm]{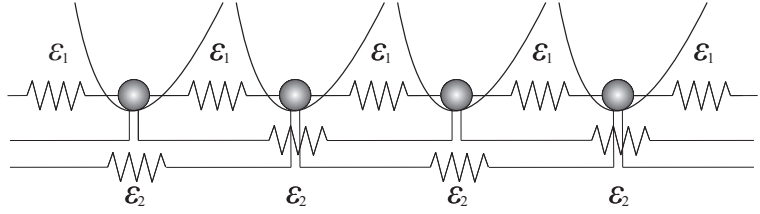}
	\caption{The $r=2$ Klein-Gordon chain with next nearest neighbor
interactions}
	\label{fig:KGchain2}
\end{figure}

The picture {\it radically} changes when the chain involves interactions with range longer than mere nearest neighbors. The range parameter
$r$ will be used to indicate the
interaction length between the oscillators of the chain. So, for the classic
nearest neighbor chain the range is $r=1$ as shown in fig.~\ref{fig:KGchain} 
while for
the next nearest neighbor (NNN) chain the range is
$r=2$ as illustrated in fig.~\ref{fig:KGchain2} etc. 
The coupling force between the
oscillators of the chain is linear and the coupling
constants $\e_i, i=1\ldots r$ are not, in general, equal.

The Hamiltonian of a 1D KG chain with long range interactions is:
\beq H=\sum_{i=-\infty}^{\infty} [\frac{p_i^2}{2} +V(x_i)] +\frac{1}{2}\sum_{i=-\infty}^{\infty}\sum_{j=1}^r\e_j\left(x_{i}-x_{i+j}\right)^2\label{ham_gen}\eeq
which leads to the equations of motion
$$\ddot{x_i}=-V'(x_i)+\sum_{j=1}^r\e_j(x_{i-j}-2x_i+x_{i+j})$$

\subsection{Persistence of multibreathers}

Let $\e_j=k_j\e$, with $k_1=1$, then the Hamiltonian (\ref{ham_gen}) becomes
\beq H=H_0+\e H_1=\sum_{i=-\infty}^{\infty} [\frac{p_i^2}{2} +V(x_i)] +\frac{\e}{2}\sum_{i=-\infty}^{\infty}\sum_{j=1}^r k_j\left(x_{i}-x_{i+j}\right)^2\label{ham_long}\eeq

Now, since the Hamiltonian is written in the form $H_0+\e\avh$ the persistence
conditions (\ref{gen_per}) can be used.
If we consider again $n+1$ ``central'' oscillators and $x_i=\sum_{m=0}^\infty A_m\cos(mw_i)$, we get for this case
\beq\avh=-\frac{1}{2}\sum_{m=1}^\infty \sum_{j=1}^r\sum_{s=1}^{n-j+1}A_m^2k_j\cos(m\sum_{l=0}^{j-1}\phi_{s+l}).\label{h1_gen}\eeq
Note that, in the above we considered $r\leqslant n$ since any interaction of
oscillators with $r>n$ does not affect the calculations, which are performed
in the anti-continuum limit. So, if one considers $r>n$ then for the
calculations in this section it would be equivalent to the choice of $r=n$. By differentiating Eq. \ref{h1_gen} with respect to $\phi_i$ we get
\beq\frac{\pa\avh}{\pa\phi_i}=0\Rightarrow\sum_{m=1}^\infty \sum_{p=1}^r\sum_{s=z_1}^{z_2}mA_m^2k_p\sin(m\sum_{l=0}^{p-1}\phi_{s+l})=0,\label{per_r}\eeq
or, by taking into consideration the definition of (\ref{per_r1}),
\beq\sum_{p=1}^r\sum_{s=z_1}^{z_2}k_p M(\sum_{l=0}^{p-1}\phi_{s+l})=0,\label{per_r_m}\eeq
where $z_1=\mathrm{max}(1,i-p+1)$ and
$\ds z_2=\left\{\begin{array}{ccc}i&\mathrm{for}&i+p-1\leqslant n\\ n-p+1 &\mathrm{for}& i+p-1>n \end{array}\right.$.

Eqs. \ref{h1_gen} and \ref{per_r} (or \ref{per_r_m}) may be seem cumbersome to handle, they are much easier to use in the concrete examples that will follow in the next sections.

\subsection{Stability of multibreathers}
As we have already mentioned the characteristic exponents of the multibreather provided by the persistence conditions (\ref{per_r}) are given, to leading order of approximation, by (\ref{sz}), i.e.
$$\sigma_{\pm i}=\pm\sqrt{\ds-\e\frac{\pa \w}{\pa J}{\chi_z}_i},\quad i=1\ldots n, $$
where ${\chi_z}_i$ are the eigenvalues of ${\bf Z}$ with
\beq{\bf Z}={\bf  B_1\cdot L}=\frac{\pa^2\avh}{\pa\phi_i\pa\phi_j}\cdot\left(\begin{array}{ccccc}
2&-1&0 & & \\
-1&2&-1&0 & \\
 &\ddots&\ddots&\ddots & \\
 & 0 &-1&2&-1\\
 &  &0 & -1&2
\end{array}\right),\quad i,j=1\ldots n.\label{z}\eeq

For linear stability we need all the characteristic exponents to be purely imaginary. So, if $P=\e\frac{\pa \w}{\pa J}<0$ we need all the eigenvalues of $\bf Z$ to be negative, while if $P=\e\frac{\pa \w}{\pa J}>0$ we need all the eigenvalues of $\bf Z$ to be positive.\\[20pt]

Without loss of generality we can consider $i\lesq j$, since the ${\bf  B_1}$ matrix is symmetric. Let $d=j-i+1$, then the general form of $\frac{\pa^2\avh}{\pa\phi_i\pa\phi_j}$ is
\beq\frac{\pa^2\avh}{\pa\phi_i\pa\phi_j}=\left\{\begin{array}{cl}
0 &\quad \mathrm{if}\quad d>r\\[8pt]
\ds\frac{1}{2}\sum_{m=1}^\infty \sum_{p=d}^r\sum_{s=z_1}^{z_2}m^2A_m^2k_p\cos(m\sum_{l=0}^{p-1}\phi_{s+l}) &\quad \mathrm{if}\quad d\lesq r
\end{array}\right.
,\label{b1}\eeq
or, by taking under consideration the definition of (\ref{eq:f}),

\beq\frac{\pa^2\avh}{\pa\phi_i\pa\phi_j}=\left\{\begin{array}{cl}
0 &\quad \mathrm{if}\quad d>r\\[8pt]
\ds\sum_{p=d}^r\sum_{s=z_1}^{z_2}k_p f(\sum_{l=0}^{p-1}\phi_{s+l}) &\quad \mathrm{if}\quad d\lesq r
\end{array}\right.
.\label{b1_f}\eeq

Now $z_1$ is given by $z_1=\mathrm{max}(1,i-p+d)$, while $z_2$ is still given by
$\ds z_2=\left\{\begin{array}{ccc}i&\mathrm{for}&i+p-1\leqslant n\\ n-p+1 &\mathrm{for}& i+p-1>n \end{array}\right. . $

In order to demonstrate the use of the results of this section, in what follows, we will examine some particular cases.

\section{3-site breathers with $r=2$ }

\subsection{The $\e_1=\e_2=\e$ case}
\subsubsection{Persistence of multibreathers}
The simplest case to check the effect of long range interactions is the one of 3 central oscilators (i.e. $n=2$) and range $r=2$. As we have already mentioned, any range $r>2$ would not affect our calculations. First we will check the case $k_1=k_2=1\Rightarrow\e_1=\e_2=\e$.

\begin{figure}
	\centering
		\includegraphics[width=6cm]{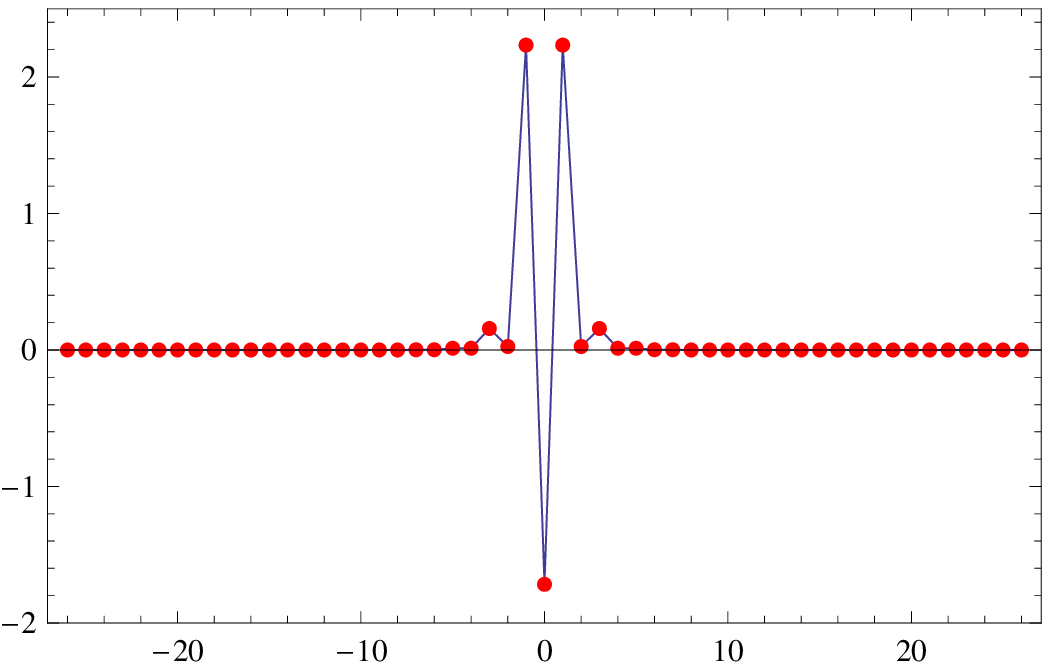}\hspace{1cm}\includegraphics[width=6cm]{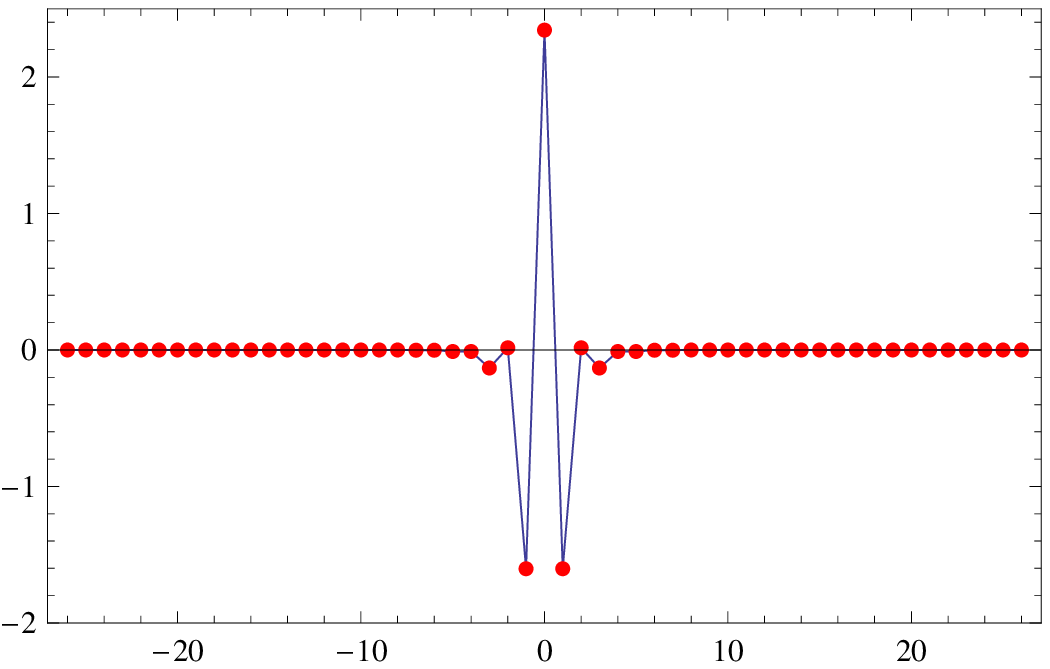}
	\caption{Two snapshots of a 3-site ($n=2$), anti-phase ($\phi_1=\phi_2=\pi$) multibreather in a range $r=2$ Klein-Gordon chain with $\e_1=\e_2=0.02$ and frequency $\w=2\pi/7$. See also video1 in \cite{videolink}.}
	\label{fig:out3}
\end{figure}

\begin{figure}
	\centering
		\includegraphics[width=5cm]{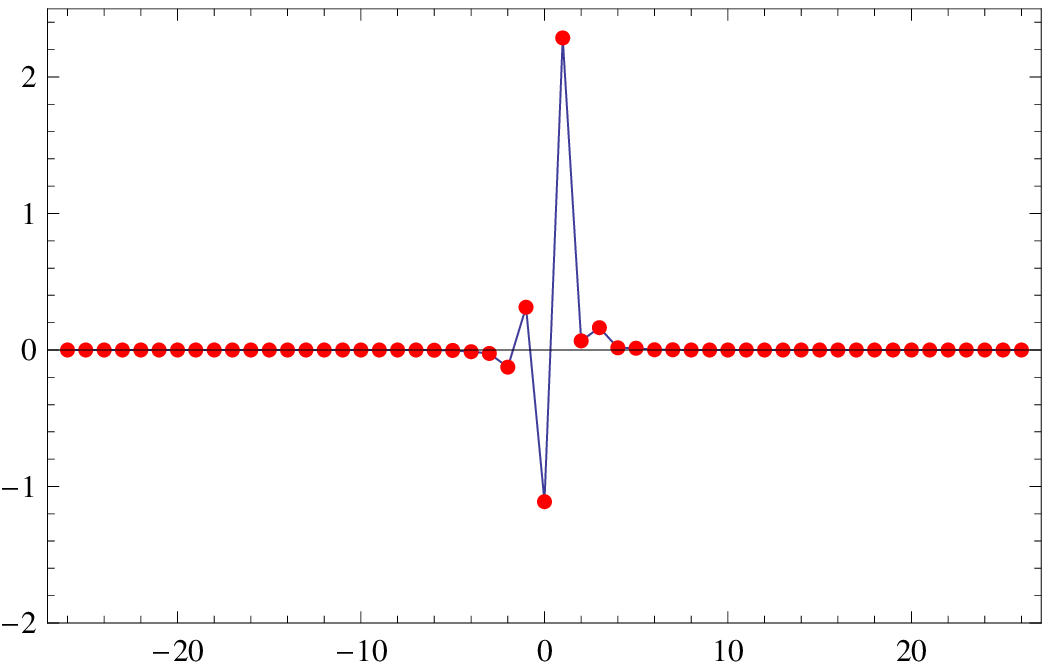}\hspace{0.5cm}\includegraphics[width=5cm]{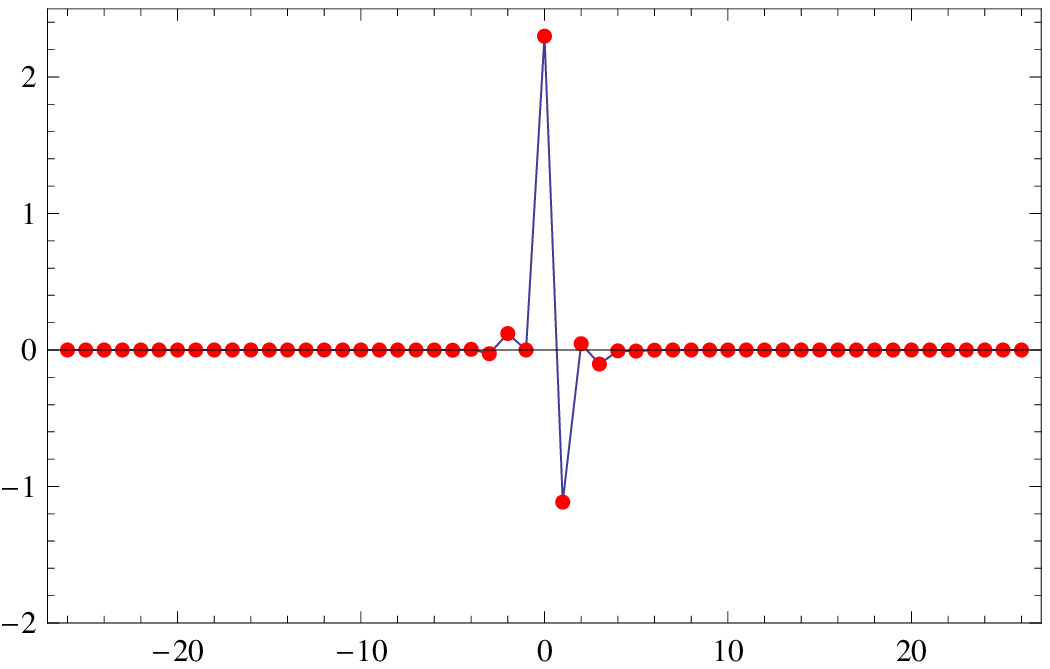}\hspace{0.5cm}\includegraphics[width=5cm]{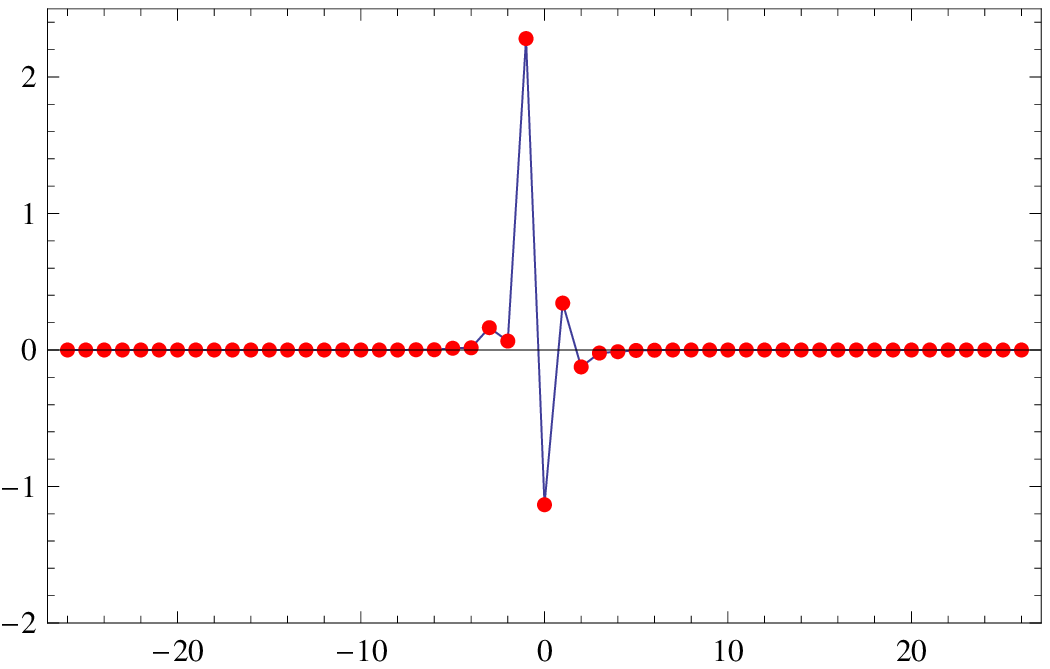}
	\caption{Three snapshots of a 3-site ($n=2$), phase-shift ($\phi_1=\phi_2\neq0,\pi$) multibreather in a range $r=2$ Klein-Gordon chain with $\e_1=\e_2=0.02$ and frequency $\w=2\pi/7$. See also video3 in \cite{videolink}.}
	\label{fig:vortex3}
\end{figure}

In this case the Hamiltonian (\ref{ham_long}) reads
$$H=H_0+\e H_1=\sum_{i=-\infty}^{\infty} \frac{1}{2}p_i^2 +V(x_i) +\frac{\e}{2}\sum_{i=-\infty}^{\infty}\left[(x_i-x_{i+1})^2+(x_i-x_{i+2})^2\right].$$
Since we consider a 3-site breather, $H_1$ becomes at the anti-continuum limit
$$H_1=x_1^2+x_2^2+x_3^2+(x_1-x_2)^2+(x_1-x_3)^2+(x_2-x_3)^2$$
and
$$\avh=-\frac{1}{2}\sum_{m=1}^\infty A_m^2\left\{\cos(m\phi_1)+\cos(m\phi_2)+\cos[m(\phi_1+\phi_2)]\right\},$$
according also to (\ref{h1_gen}), for $n=2$ and $r=2$.
The persistence conditions (\ref{per_r}) become
\beq\frac{\pa \avh}{\pa \phi_i}=0\Rightarrow\sum_{m=1}^\infty mA_m^2\left\{\sin(m\phi_i)+\sin[m(\phi_1+\phi_2)]\right\}=0,\quad\text{for}\quad i=1,2\label{eq:con2}\eeq
or, by taking into consideration the definition of $M(\phi)$ in (\ref{per_r1}),
\beq M(\phi_i)+M(\phi_1+\phi_2)=0,\quad \text{for}\quad i=1,2.\label{eq:con2_m}\eeq
This equation, in addition to the standard solutions
$$\phi_i=0, \pi,$$
provides also the solutions
$$\phi_1=\phi_2=2\pi/3, \phi_1=\phi_2=4\pi/3.$$
The multibreather solutions with $\phi_i\neq0$ are called {\it phase-shift multibreathers} or phase-shift breathers. The anti-phase and phase-shift configurations are depicted in figs. \ref{fig:out3} and \ref{fig:vortex3}. For a better visualization one can also refer to videos 1 and 3 in \cite{videolink} (video2 shows an in-phase configuration).

In order to produce these figures (and videos) 
we used the on-site potential $V(x)=\ds\frac{x^2}{2}-0.15\frac{x^3}{3}-0.05\frac{x^4}{4}$ and initial conditions which correspond to motion with period $T=7$ and frequency $\w=2\pi/T=2\pi/7\simeq0.8976$. The same potential is used for every numerical calculation throughout this work, although it
is straightforward to apply the relevant notions to arbitrary potentials of
the Klein-Gordon type.

\noindent {\bf Remark 1:} The persistence conditions (\ref{eq:con2_m}) provide 2 equations.
\beq\begin{array}{l}
M(\phi_1)+M(\phi_1+\phi_2)=0\\
M(\phi_2)+M(\phi_1+\phi_2)=0.
\end{array}\label{eq:con22}\eeq
By substraction of equations (\ref{eq:con22}) we get
\beq M(\phi_1)=M(\phi_2) \label{eq:con23}\eeq
which has, besides the trivial solutions $\phi_i=0,\pi$, two other obvious solutions: $\phi_1=\phi_2$ and $\phi_1+\phi_2=\pi$ for $0\lesq\phi_i\lesq 2\pi$. The last solution does not provide any new information because by substituting this into equations (\ref{eq:con22}) we get $M(\phi_1)=M(\phi_2)=0$, which
, as it is shown in \cite{kouk12}, 
only possesses the $\phi_i=0, \pi$ solutions.
But the $\phi_1=\phi_2=\phi$ solution can reduce the two equations (\ref{eq:con22}) into equation (\ref{eq:con2_m_red})
\beq M(\phi)+M(2\phi)=0.\label{eq:con2_m_red}\eeq
\noindent {\bf Remark 2:} Our numerical computations strongly suggest that  for all the phase-shift solutions it is $\phi_1=\phi_2$, yet a rigorous proof of this fact is still an open problem. So, equation (\ref{eq:con2_m_red}) can be used in order to calculate all the solutions of the persistence conditions (\ref{eq:con2_m}), except for the mixed one \{$\phi_1=0,\ \phi_2=\pi$\} (or equivalently \{$\phi_1=\pi,\ \phi_2=0$\}).\\[4pt]

\noindent {\bf Remark 3:} In the case under consideration ($n=2$, $r=2$, $k_i=1$), all the available solutions correspond to $\phi_i$'s which make
 each of the terms of the sum vanish in (\ref{eq:con2}) which obviously provides a zero total.

\noindent {\bf Remark 4:} The case under consideration
is equivalent to the $3$-site breathers on a hexagonal lattice which has
already been studied in \cite{koukmac,kouketal}. It can be effectively
considered as a one-dimensional realization of such a lattice.
In that context, the phase-shift multibreathers can be alternatively
thought as ``discrete vortices'', as they are solutions which complete
a phase rotation by $2 \pi$, as one traverses a discrete contour
(which consists of the relevant triangle of sites).

\noindent {\bf Remark 5:} As an aside, it should be mentioned that
an additional
motivation for the consideration of such next-nearest neighbor
interactions stems from the consideration of zigzag arrays, similar
to the waveguide arrays proposed theoretically in the context
of nonlinear optics (and hence in the realm of the DNLS equation)
in~\cite{efremdnc}.

\noindent {\bf Remark 6:} The stability of the above mentioned breathers will be discussed at the end of the next section as a special case of the more
general unequal coupling one.

\subsection{The $\e_1\neq\e_2$ case}
\subsubsection{Persistence of multibreathers}
Although the $\e_1=\e_2$ case is the easiest and allows us to perform some analytic calculations as well, the natural consideration for the case of 
next-nearest neighbors is the one with $\e_1\neq\e_2$. Intuitive
physical considerations suggest to enforce $\e_1\gesq\e_2$ (considering coupling force decreasing with the distance between the oscillators) but there are configurations (like the zigzag one) which may also justify settings with $\e_1 < \e_2$~\cite{efremdnc}. Let $k_1=1$ and $k_2=k$ or, $\e_1=\e$ and $\e_2=k\e$. In this case, the Hamiltonian (\ref{ham_long}) reads
$$H=H_0+\e H_1=\sum_{i=-\infty}^{\infty} \frac{1}{2}p_i^2 +V(x_i) +\frac{\e}{2}\sum_{i=-\infty}^{\infty}\left[(x_i-x_{i-1})^2+k(x_i-x_{i-2})^2\right]. $$
Since we consider a $3$-site breather ($n=2$) we have only two independent $\phi_i$'s in the anti-continuum limit and by (\ref{h1_gen}) we get
$$\avh=-\frac{1}{2}\sum_{m=1}^\infty A_m^2\left\{\cos(m\phi_1)+\cos(m\phi_2)+k\cos[m(\phi_1+\phi_2)]\right\}.$$
This leads to the persistence conditions:
\beq\frac{\pa \avh}{\pa \phi_i}=0\Rightarrow\sum_{m=1}^\infty mA_m^2\left\{\sin(m\phi_i)+k\sin[m(\phi_1+\phi_2)]\right\}=0\equiv M(\phi_i)+k M(\phi_1+\phi_2)=0\quad \text{for}\quad i=1,2.\label{per_r2_n2_gen}\eeq

\noindent {\bf Remark:} By using the same arguments as in the previous section, if we consider $\phi_1=\phi_2$, we get from (\ref{per_r2_n2_gen}),
\beq\sum_{m=1}^\infty mA_m^2\left[\sin(m\phi)+k\sin(2m\phi)\right]=0\equiv M(\phi)+k M(2\phi)=0.\label{per_r2_n2_gen_simpl}\eeq

So, one could use (\ref{per_r2_n2_gen_simpl}) instead of (\ref{per_r2_n2_gen}) as the relevant persistence condition in order to calculate all the solutions of (\ref{per_r2_n2_gen}) except of the mixed one \{$\phi_1=0,\ \phi_2=\pi$\} (or equivalently \{$\phi_1=\pi,\ \phi_2=0$\}).

\begin{figure}
	\centering
		\includegraphics[width=10cm]{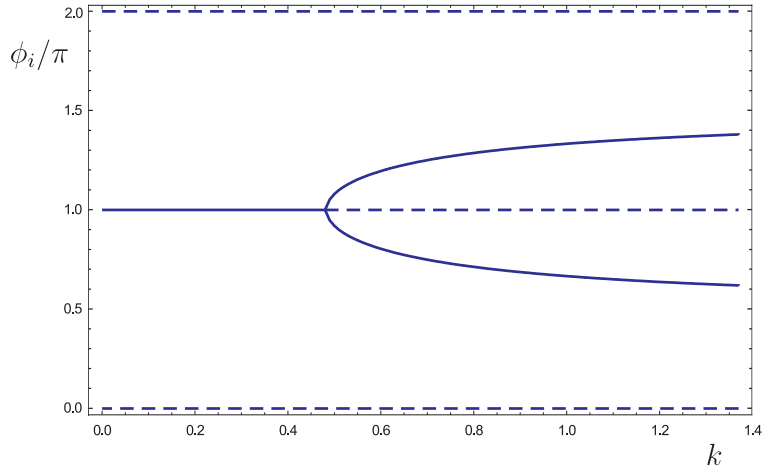}
		\caption{The bifurcation diagram for a 3-site ($n=2$) breather in a $r=2$ Klein-Gordon chain with $k=\e_2/\e_1$. A pitchfork bifurcation occurs for $k=k_{cr}=0.48286$. The curve is calculated using two methods. The first method is to use the full model, calculate the multibreather solutions with $\e=0.001$ and depict them as well as their stability. The second method is to solve numerically (\ref{per_r2_n2_gen}) and check when solutions with $\phi_i\neq0,\pi$ appear. The curves produced with the two methods practically coincide (i.e., no
difference is discernible at the scale of the plot). The various families that appear here are depicted in more detail in Fig. \ref{pitchfork_parts}.}
		\label{pitchfork}
\end{figure}

In the $k=1$ ($\e_1=\e_2$) case, one could make a choice of $\phi_1=\phi_2=2\pi/3\ \mathrm{or}\ 4\pi/3$ in order to have $\left\{\sin(m\phi_i)+\sin[m(\phi_1+\phi_2)]\right\}=0$ $\forall m$, so that the total sum in (\ref{eq:con2}) would vanish also. This is not possible in the $k\neq 1$ case. So, one may be led to believe that this is an
isolated solution and that possibly there are no other solutions than $\phi=0$ and $\phi=\pi$ in this case. However, it
instead turns out that there can be other solutions also which can be
calculated numerically  for $k\neq1$. In fact, there is a critical
value $k_{cr}=0.48286$ of $k$ where a pitchfork bifurcation occurs
(fig.~\ref{pitchfork}). For values $k<k_{cr}$ the only solutions
Eq. (\ref{per_r2_n2_gen}) [or (\ref{per_r2_n2_gen_simpl})] has are the
trivial ones $\phi_i=0, \pi$. For $k>k_{cr}$, i.e., past the
supercritical pitchfork bifurcation point,  other solutions
appear with $\phi_i\neq0, \pi$ (phase-shift breathers) as is shown in fig.~\ref{pitchfork}.

The bifurcation curve has been calculated in two ways. Firstly by
numerically modeling the full system and secondly by numerically solving
the transcendental existence conditions (\ref{per_r2_n2_gen}) using a small value of $\e=0.001$. The two curves
practically coincide, which illustrates the remarkable accuracy of the theory
in the vicinity of the anti-continuum limit.

\begin{figure}
	\centering
		\includegraphics[width=5cm]{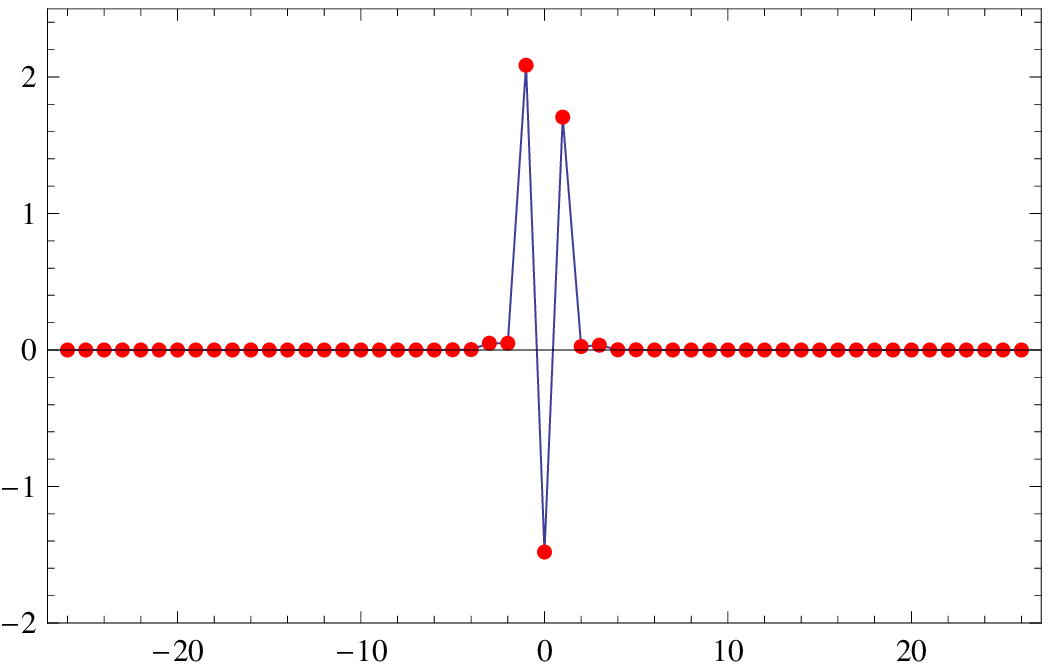}\hspace{0.5cm}\includegraphics[width=5cm]{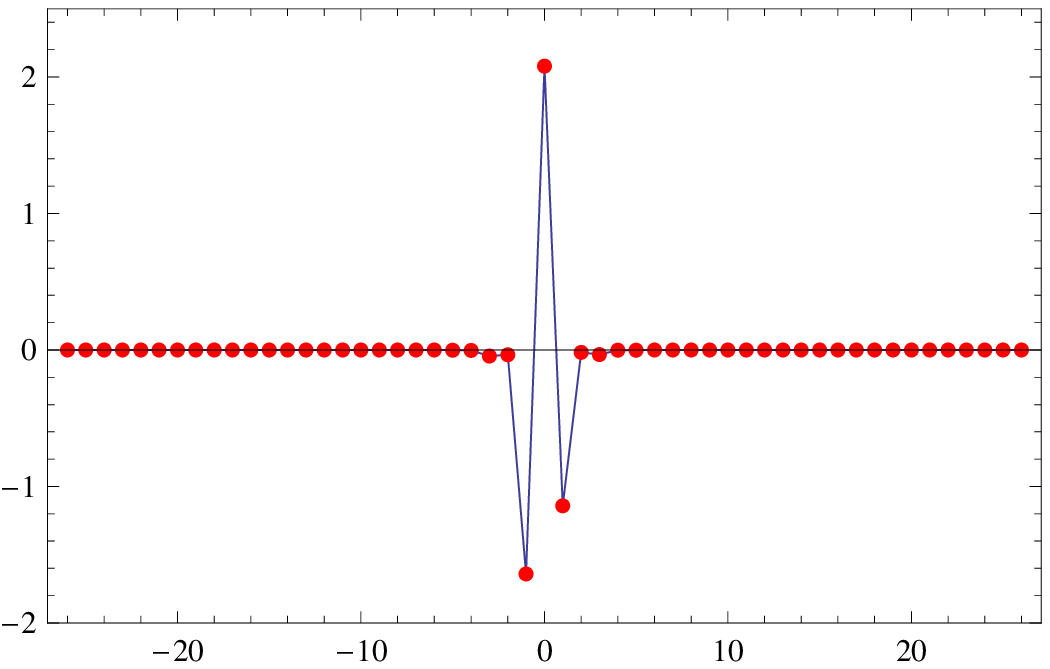}\hspace{0.5cm}\includegraphics[width=5cm]{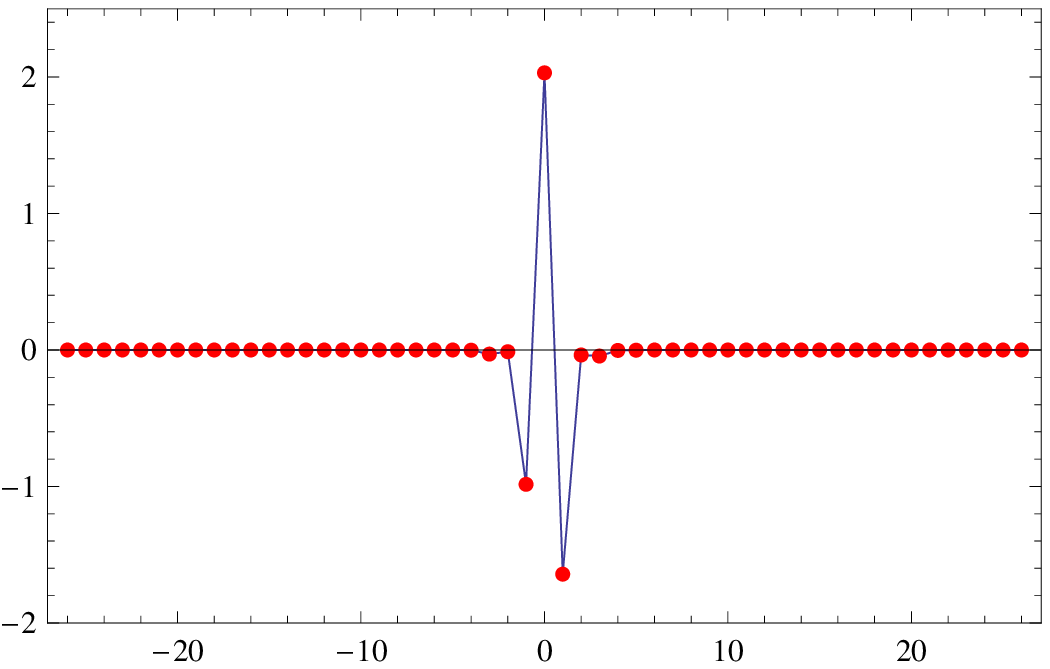}\hspace{0.5cm}
  	\caption{Snapshots of a phase-shift 3-site breather for $\e_1=0.02$ and $\e_2=k_2\e_1=0.56\e_1$ and frequency $\w=2\pi/7$. See also video4 in \cite{videolink}.}
	\label{fig:e1neqe2}
\end{figure}

A phase-shift breather with
$k=0.54$ is depicted in fig.~\ref{fig:e1neqe2}. For a better visualization
of this breather one can also see video4 in \cite{videolink}.

\begin{figure}
\begin{tabular}{ccc}
\includegraphics[width=6cm]{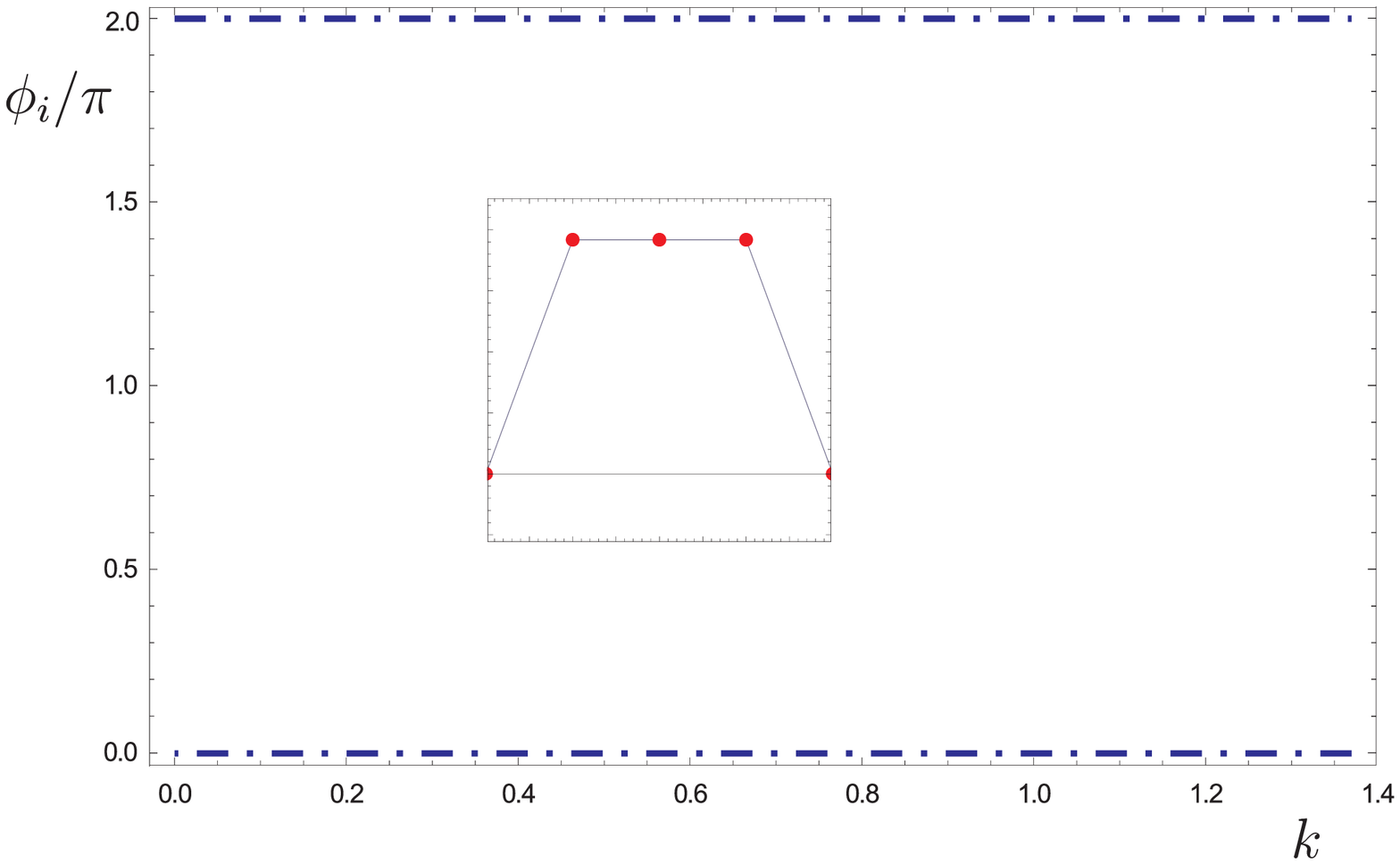}&\includegraphics[width=6cm]{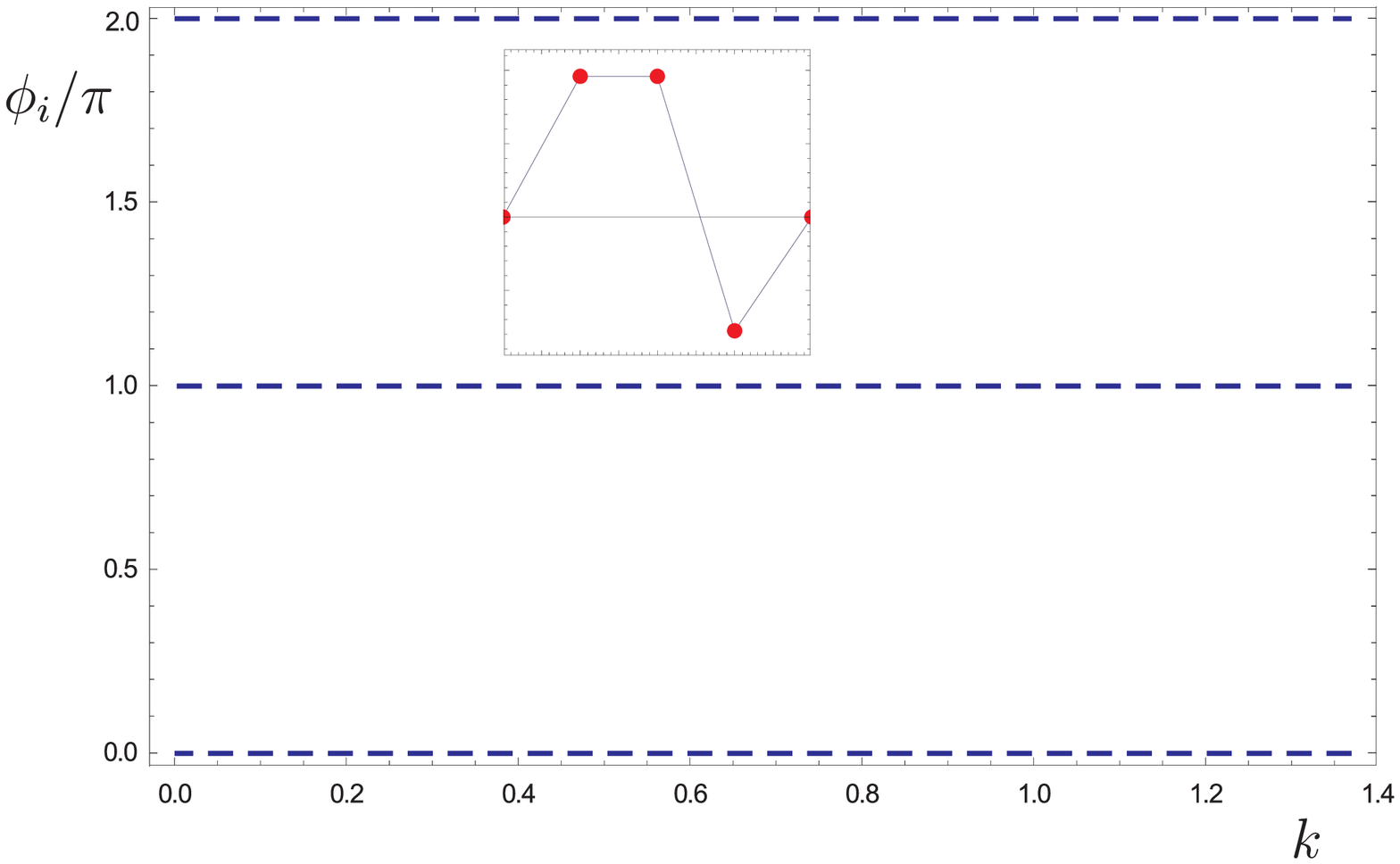}&\includegraphics[width=6cm]{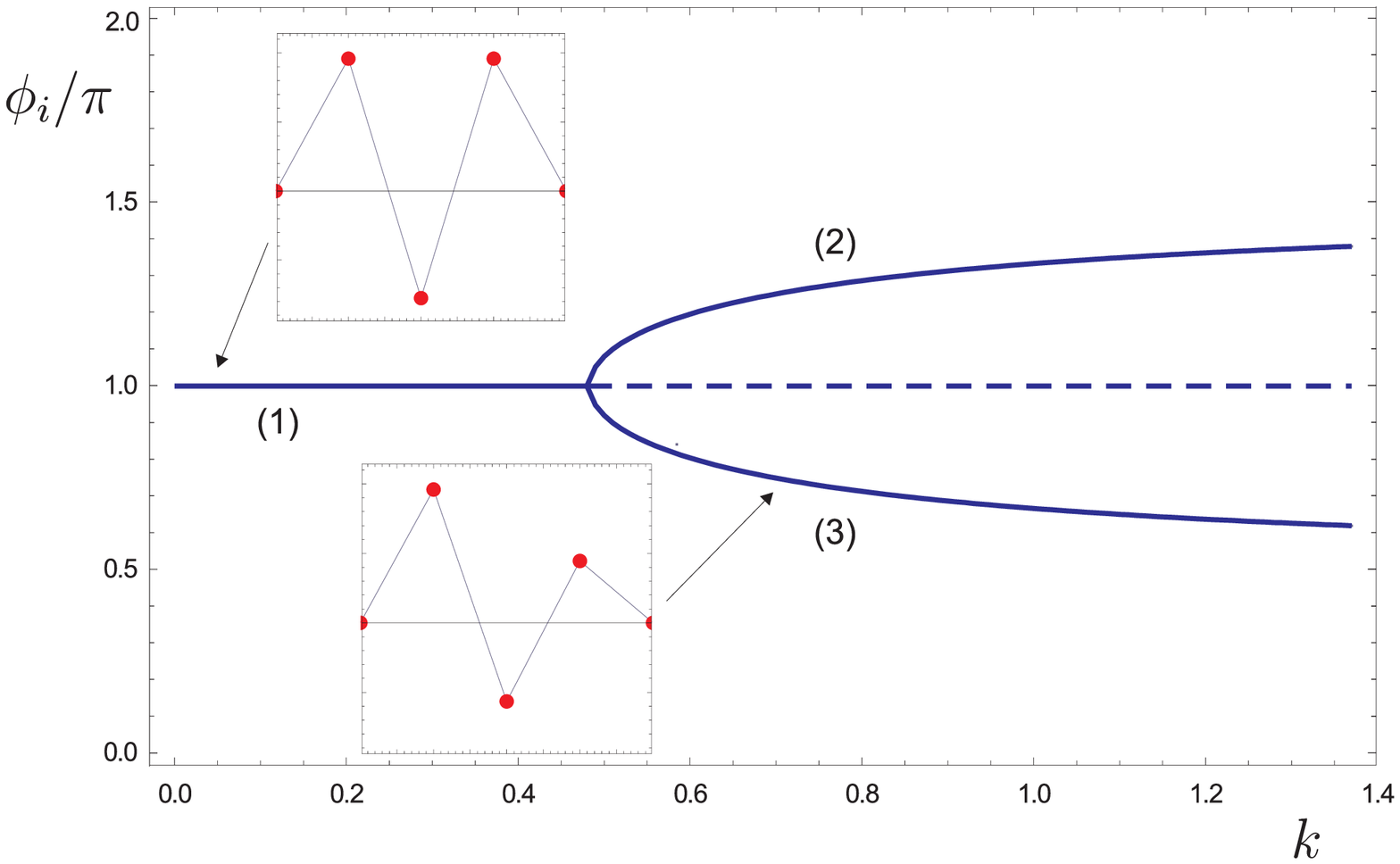}\\
(a)&(b)&(c)
\end{tabular}
\caption{The various families that constitute Fig.~\ref{pitchfork} are depicted. Portraits of the configuration of the central oscillators for the different multibreather families are shown as insets in the corresponding diagrams. 
The line type in the figures depend in the number of positive $\xz$ 
(equivalently, the number of real eigenvalue pairs for
$P\equiv\ds\e\frac{\pa \w}{\pa J}<0$) the corresponding family possesses: no positive $\xz$ corresponds to solid line, 1 positive $\xz$ corresponds to dashed line and 2 to dashed-dotted line. In {\bf (a)} the in-phase configuration is depicted \{$\phi_1=\phi_2=0\ (\text{or}\ 2\pi)$\} which possesses 2 positive $\chi_z$. In {\bf (b)} the mixed configuration is shown \{$\phi_1=0,\ \phi_2=\pi$\} which possesses 1 positive and 1 negative $\chi_z$. In {\bf (c)} two families are shown. The first is the anti-phase one~\{$\phi_1=\phi_2=\pi$\}. It has 2 negative $\chi_z$ until $k<k_{cr}=0.48286$ while it has 1 positive and 1 negative $\chi_z$ for $k>k_{cr}$. At this point the anti-phase family bifurcates to provide the phase-shift configuration \{$\phi_1=\phi_2\neq0, \pi$\}. This family is represented by $\phi_1=\phi_2=(2)$ or $\phi_1=\phi_2=(3)$ 
and has no positive (2 negative) $\chi_z$. All of these together are depicted in Fig.~\ref{pitchfork}. When two line segments coincide the more dense is shown.}
\label{pitchfork_parts}
\end{figure}

\subsubsection{Stability of multibreathers}

By using the previously developed theory, we can calculate the characteristic exponents of the various configurations of 3-breathers in this lattice setting. The characteristic exponents of the specific solutions are given to first order of approximation by (\ref{sz}) as

$$\sigma_{\pm i}=\pm\sqrt{-\e\frac{\pa \w}{\pa J}{\chi_z}_i},$$
where ${\chi_z}_i$ are the eigenvalues of the matrix ${\bf Z}$ defined in (\ref{z}).

In the case under consideration of 3-site ($n=2$) breathers with range $r=2$ we have, also from (\ref{b1_f}),
$$\frac{\pa^2\avh}{\pa\phi_i\pa\phi_j}=\left(\begin{array}{cc}f(\phi_1)+k f(\phi_1+\phi_2)&k f(\phi_1+\phi_2)\\k f(\phi_1+\phi_2)& f(\phi_2)+k f(\phi_1+\phi_2)\end{array}\right)
\quad\mathrm{and}\quad
{\bf L}=\left(\begin{array}{cc}2&-1\\-1&2\end{array}\right).$$
So, we get from (\ref{z}),
$${\bf Z}=\frac{\pa^2\avh}{\pa\phi_i\pa\phi_j}\cdot{\bf L}=\left(\begin{array}{cc}2f_1+k f_{1+2}&k f_{1+2}-f_1\\k f_{1+2}-f_2& 2f_2+k f_{1+2}\end{array}\right),$$
where the function $f(\phi)$ is defined as in (\ref{eq:f}), and $f_{1+2}\equiv f(\phi_1+\phi_2)$, while $f_i=f(\phi_i)$ for $i=1,2$ .

For linear stability it is required that all of the characteristic
exponents be purely imaginary. So, the stability is determined by the sign of ${\chi_z}_i$.

In particular, we
check the configurations that can appear in this case.
\begin{itemize}
\item $\phi_1=\phi_2$.  This is the general case and includes the in-phase \{$\phi_1=\phi_2=0$\}, the out-of-phase \{$\phi_1=\phi_2=\pi$\} and phase-shift configurations \{$\phi_1=\phi_2\neq0, \pi$\}. The corresponding eigenvalues $\chi_z$ are ${\chi_z}_1=3f_\phi$ and ${\chi_z}_2=f_\phi+2k f_{2\phi}$.\\
\item $\phi_1=0, \phi_2=\pi$. This is the only solution with $\phi_1\neq\phi_2$. For this case it is ${\chi_z}_{1,2}=f_0+(1+k)f_{\pi}\pm\sqrt{f_{0}^2-(1+k)f_0 f_{\pi}+f_{\pi}^2(1-k+k^2)}$.\\
\end{itemize}

\noindent{\bf Remark:} We have that $f_0>0$ as a direct consequence of the definition (\ref{eq:f}) of $f(\phi)$. On the other hand it is $f_\pi<0$. This can be rigorously proven (\cite{KK09} Lemma 3)
but it can also be intuitively 
understood by the definition (\ref{eq:f}) of $f(\phi)$ and the fact that the first term of the Fourier expansion of $x(w)$ (\ref{xdevel}) is the dominant one. Using the same arguments we can conclude that $|f_0|>|f_\pi|$. So, we can immediately conclude that in the in-phase \{$\phi_1=\phi_2=0$\} configuration it is ${\chi_z}_i>0$, while in the mixed \{$\phi_1=0,\ \phi_2=\pi$\} configuration it is ${\chi_z}_{1}>0$, ${\chi_z}_{2}<0$.

On the other hand for the anti-phase \{$\phi_1=\phi_2=\pi$\} configuration, the formulas for the ${\chi_z}_i$ read
\beq{\chi_z}_1=3f_\pi\quad \text{and}\quad {\chi_z}_2=f_\pi+2k f_{0}.\label{eq:xi_r2}\eeq
The ${\chi_z}_1$ eigenvalue is always negative while the sign of the ${\chi_z}_2$ depends on the value of $k$. This can provide us with a criterion about the value of $k_{cr}$ where the bifurcation occurs, since, at this point ${\chi_z}_2$ changes sign. So, by (\ref{eq:xi_r2}) we get $\xz_2=0\Rightarrow k_{cr}=\ds-\frac{f_\pi}{2f_0}$. 

The values of $f_0$ and $f_\pi$ depend on the particular on-site potential as well as on the frequency we examine, so, the value $k_{cr}$ is not fixed. But, if we consider breathers with relatively low amplitude, which amounts to 
the breather frequency $\w$ being close to the phonon frequency $\w_p$, a rough estimation of $k_{cr}$ can be made. In such a  case, 
the nonlinear character of the system is not fully revealed yet which 
means that the $A_1$ term in the development (\ref{xdevel}) is by far the 
most dominant one. This results to $|f_0|\simeq|f_\pi|$ and 
consequently $k_{cr}\simeq0.5$.

\begin{figure}[ht]
	\centering
		\includegraphics[width=8cm]{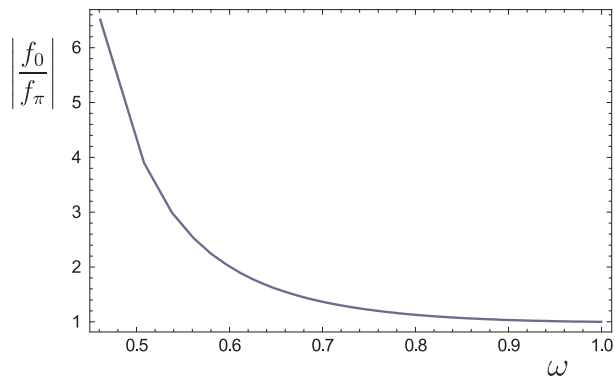}
	\caption{Dependence of the $\left| f_0/f_\pi \right|$ ratio with respect to the frequency $\w$ of the breather.}
	\label{fig:f0_fpi_w}
\end{figure}

In order to check our estimation, we perform some numerical calculations for the lattice with potential $V(x)=\ds\frac{x^2}{2}-0.15\frac{x^3}{3}-0.05\frac{x^4}{4}$ which we use throughout this work, considering a motion with $\w=2\pi/7\simeq0.8976$. For this frequency, it is $\left| f_0/f_\pi \right|\simeq1$, as 
can be seen in Fig. \ref{fig:f0_fpi_w}, so our estimation holds. In particular, it is $f_\pi=-1.48658$, $f_0=1.53934$ and $k_{cr}=\ds-\frac{f_\pi}{2f_0}=0.48286$, which is precisely the value where the bifurcation occurs, while being very close also to the rough estimation (of $0.5$) above. Note that, as it can be seen in Fig. \ref{fig:f0_fpi_w}, if we had chosen a smaller breather frequency $\w$, our estimation would be completely mistaken, since for small values of $\w$ it is $\left| f_0/f_\pi \right|>1$.

\noindent{\bf Remarks about figs.~\ref{pitchfork} and ~\ref{pitchfork_parts}:} In figs.~\ref{pitchfork} and ~\ref{pitchfork_parts} all the multibreather families that exist in the present configuration ($n=2$, $r=2$) are shown. The multibreather families correspond to solution families of Eqs. \ref{per_r2_n2_gen}. These families are categorized by the phase differences $\phi_i$ between the successive oscillators in the anticontinuous limit. The values of $\phi_i$ in the usual families ($\phi_i=0,\pi$) are constant with increasing $k$, while in the phase-shift ($\phi_i\neq0, \pi$) families their values change with respect to $k$. 

The various solution families are represented by various line (or curve) segments in the figures. The kind of the line depends on the number 
of positive $\xz$ (i.e., of real eigenvalue pairs
for $P\equiv\ds\e\frac{\pa \w}{\pa J}<0$)
that the corresponding solution has. So, for no $\chi_z>0$ we use a solid line, for one $\chi_z>0$ we use a dashed line while for two $\chi_z>0$ we use a dashed-dotted line. Since in Fig. ~\ref{pitchfork} some of the families coincide, we separated the information in this figure into 3 panels in Fig.~\ref{pitchfork_parts}. These 3 panels together compose Fig. \ref{pitchfork}. If the segments which represent two or more distinct families of solutions coincide, the more dense is shown in the figure. In order to facilitate the visualization of 
the various families, we added insets in Figs. \ref{pitchfork_parts} 
demonstrating the profiles of (and hence illustrating 
the phase difference between) the central oscillators in the 
anticontinuum limit.  
This has as a result only solid and dashed segments to appear in Fig.~\ref{pitchfork}. The families that are depicted in the figure are:

\begin{itemize}
\item \{$\phi_1=\phi_2=0\ (\text{or}\ 2\pi)$\} (in-phase). This family is shown in Fig.~\ref{pitchfork_parts}(a) and possesses 2 positive $\chi_z$.
\item \{$\phi_1=0\ \text{or}\ \pi,\ \phi_2=\pi$\} (mixed). This family is depicted in Fig.~\ref{pitchfork_parts}(b) and possesses 1 positive and 1 negative $\chi_z$.
\item \{$\phi_1=\phi_2=\pi$\} (anti-phase). It is represented in Fig.~\ref{pitchfork_parts}(c) by $\phi_1=\phi_2=(1)$. It has no positive $\chi_z$ until $k<k_{cr}$ while it has 1 positive and 1 negative $\chi_z$ for $k>k_{cr}$. At this point the $\phi_1=\phi_2=\pi$ family becomes subject to the bifurcation
that gives rise to phase-shift multibreathers.
\item $\phi_1=\phi_2\neq0, \pi$ (phase-shift). This family is represented in Fig.~\ref{pitchfork_parts}(c) by $\phi_1=\phi_2=(2)$ or $\phi_1=\phi_2=(3)$ and has no positive $\chi_z$.
\end{itemize}
Since the stability of the multibreathers is also determined by the sign of $P\equiv\ds\e\frac{\pa \w}{\pa J}$, the above are summarized, in terms of stability of the solutions, in Table \ref{table1}.\\

\renewcommand{\arraystretch}{1.2}
\begin{table}[ht]
\begin{tabular}{|c|c|c|c|c|c|c}
\hline
 &$\hspace{0.5cm}P\hspace{0.5cm}$&$\hspace{0.7cm}k\hspace{0.7cm}$&\begin{tabular}{c}\\[-7pt] In-phase\\[-3pt] $\phi_1=\phi_2=0$\\[5pt] \end{tabular}&\begin{tabular}{c}Out-of-phase\\[-3pt] $\phi_1=\phi_2=\pi$\end{tabular}&\begin{tabular}{c} Phase-shift\\[-3pt] \mbox{$\phi_1=\phi_2\neq 0,\pi$}\end{tabular}\\
 \hline
\multirow{4}{*}{\begin{tabular}{c}Linear\\ Stability\end{tabular}}&$P<0$&$k<k_{cr}$&unstable&stable& --\\
 & $P<0$ & $k>k_{cr}$ & unstable & unstable & stable \\
 & $P>0$ & $k<k_{cr}$ & stable & unstable & -- \\
 & $P>0$ & $k>k_{cr}$ & stable & unstable & unstable \\
 \hline
\end{tabular}
\caption{Stability of the various $n=2$, $r=2$, breather configurations depending on the values of $P\equiv\frac{\pa \w}{\pa J}$ and $k$. With the dash we denote that this particular family does not exist for this range of values of $k$.}
\label{table1}
\end{table}

\noindent{\bf Stability of the various 3-site breather configurations in the $\e_1=\e_2$ case:} Using the above derived results we can conclude what it is already known from \cite{koukmac,kouketal}. i.e. for $P<0$, as long as $k<k_{cr}$ the only stable configuration is the anti-phase one, while for $k>k_{cr}$ the stable configuration is the phase-shift one (which corresponds in this case to the ``vortex'' configuration of \cite{koukmac,kouketal}). On the other hand for $P>0$ the only stable configuration is the in-phase one.

\section{4-site breathers with $r=2$}

In the next configuration we will consider four central oscillators, in order to study larger configurations, but we will keep the range to $r=2$ as a first step.

\subsection{Persistence of multibreathers}
\begin{figure}[htbp]
\includegraphics[width=10cm]{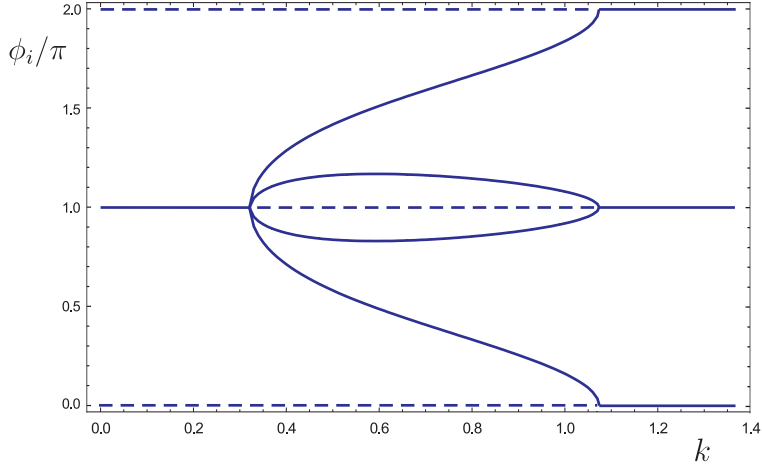}
\caption{In this diagram all the existing families of 4-site ($n=3$) breathers in a KG chain with range $r=2$ are depicted. For a more detailed view of the particular families appearing in this diagram, refer also to fig. \ref{pol_0_parts}.}
\label{fig:roots_4site_r2}
\end{figure}

We will treat the two cases $\e_1=\e_2$ and $\e_1\neq\e_2$ together, since the latter is a special case of the former  with $k_1=k_2=1$. Since we consider 4 $(n=3)$ central oscillators and range $r=2$, (\ref{h1_gen}) gives for $k_1=1$ and $k_2=k$,
$$\begin{array}{l}\ds \avh=-\frac{1}{2}\sum_{m=1}^\infty A_m^2\left\{ \cos(m\phi_1)+ \cos(m\phi_2)+ \cos(m\phi_3)+\right.\\
\left.\qquad\qquad\qquad\qquad+ k\cos[m(\phi_1+\phi_2)]+ k\cos[m(\phi_2+\phi_3)] \right\}\end{array}$$

while, the corresponding persistence conditions (\ref{per_r}) become
$$\begin{array}{ccl}
M(\phi_1)+k\, M(\phi_1+\phi_2)&=&0\\[8pt]
M(\phi_2)+k\left[ M(\phi_1+\phi_2)+M(\phi_2+\phi_3)\right]&=&0\\[8pt]
M(\phi_3)+k\, M(\phi_2+\phi_3)&=&0
\end{array}$$
which have the trivial solutions $\phi_i=0, \pi$, as well as non trivial ones, as can be seen in fig.~\ref{fig:roots_4site_r2}.

\begin{figure}[htb]
\begin{tabular}{ccc}
\includegraphics[width=7cm]{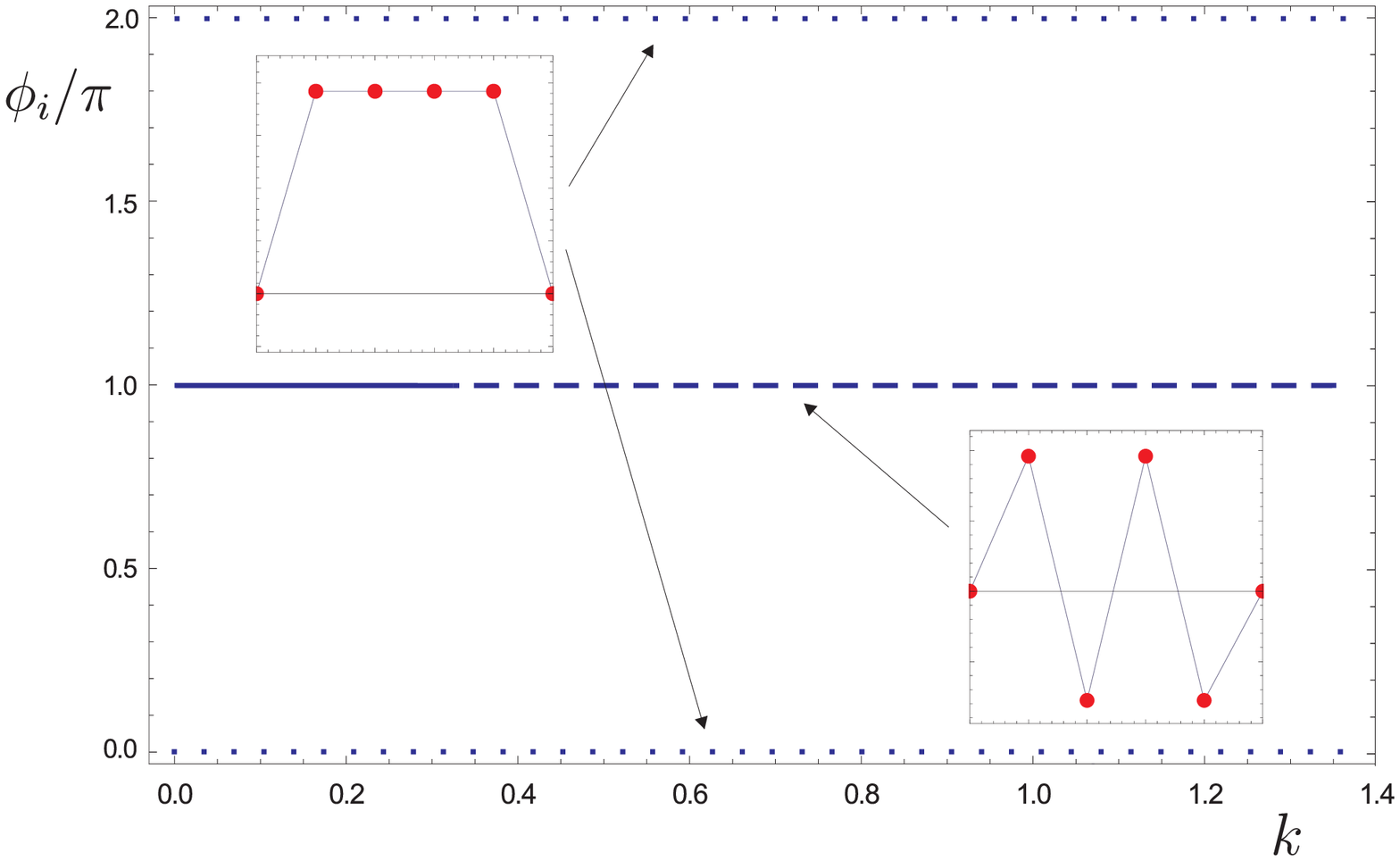}&$\hspace{0.5cm}$&\includegraphics[width=7cm]{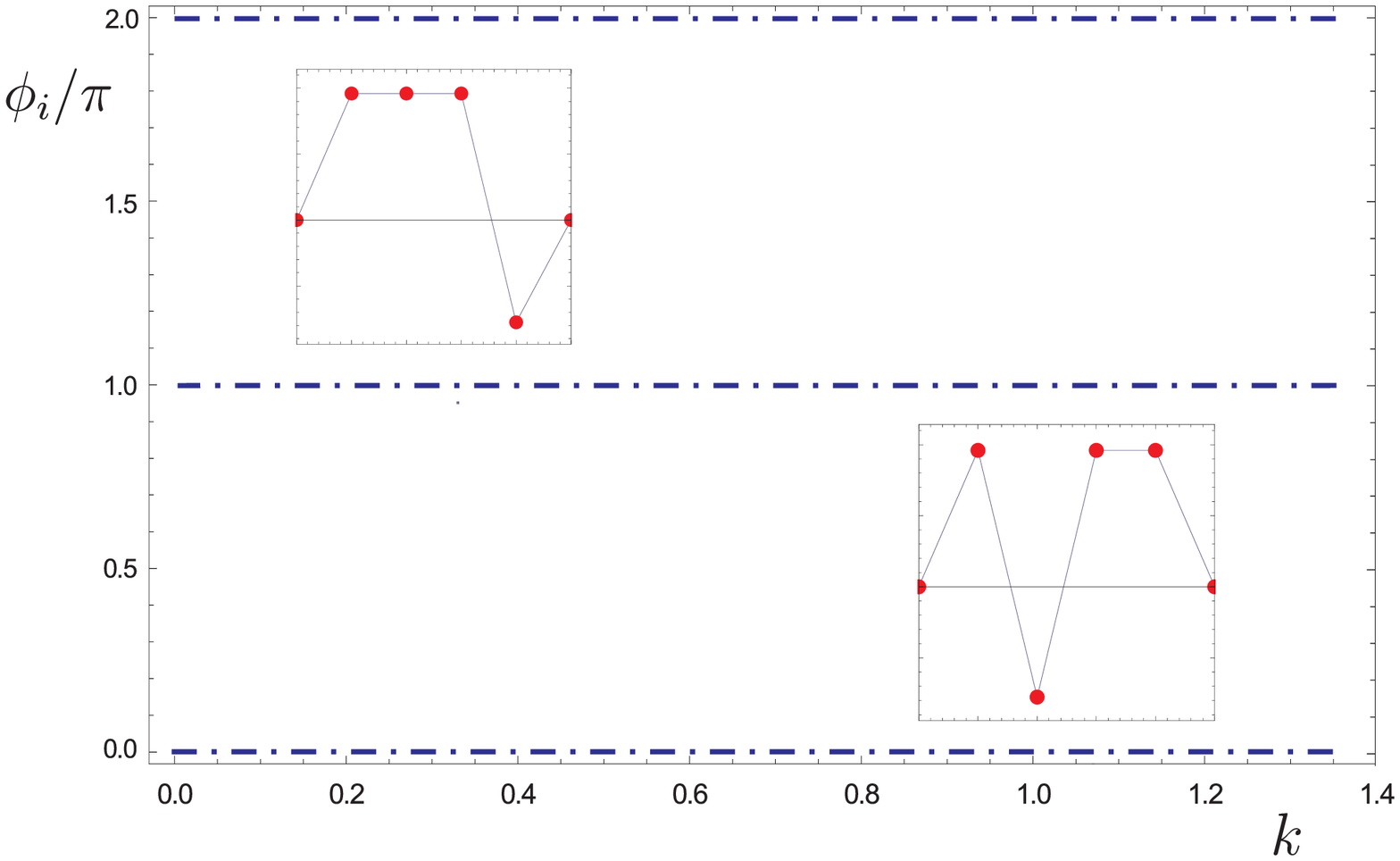}\\[-8pt]
(a)& &(b)\\[15pt]
\includegraphics[width=7cm]{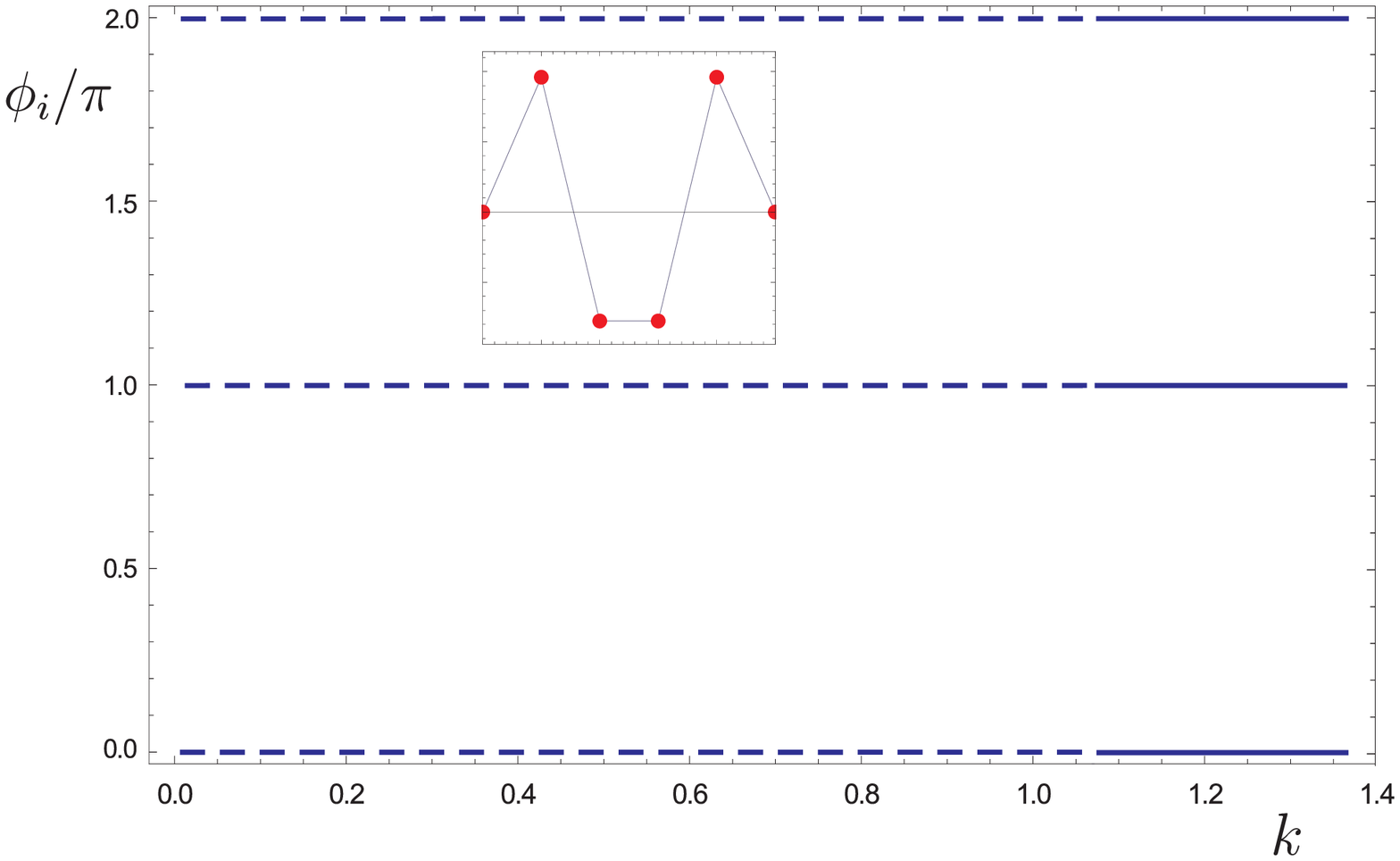}& &\includegraphics[width=7cm]{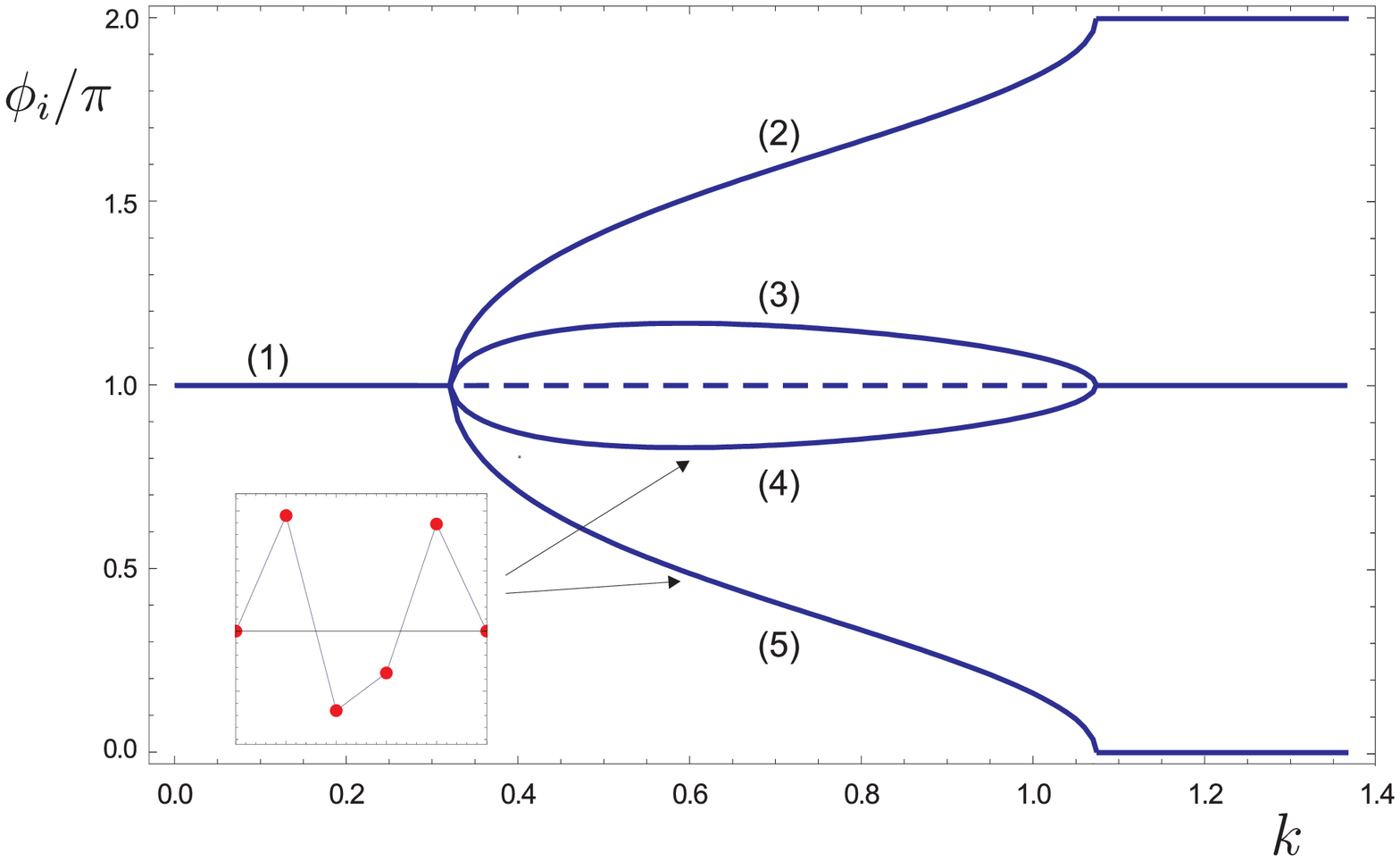}\\[-8pt]
(c)& &(d)
\end{tabular}
\caption{The various families that appear in Fig.~\ref{fig:roots_4site_r2} are depicted. Portraits of the configuration of the central oscillators for the different multibreather families are shown as insets in the 
corresponding diagrams. The line (or curve) type in the figures depends on the number of positive $\xz$ that the corresponding family possesses: no positive $\xz$ corresponds to a solid line, 1 positive $\xz$ corresponds to a dashed line, 2 to a
dash-dotted and 3 to a dotted line. In {\bf (a)} two multibreather families are depicted. The first is the in-phase one with \{$\phi_1=\phi_2=\phi_3=0\ (\text{or}\ 2\pi)$\} which possesses 3 positive $\chi_z$. The second one is the anti-phase one with \{$\phi_1=\phi_2=\phi_3=\pi$\}. It has no positive $\chi_z$ until $k<k_{cr}^{(1)}=0.3219$ while it has 1 positive $\chi_z$ for $k>k_{cr}^{(1)}$. At this point the anti-phase family
gives rise, through a supercritical pitchfork, to the the phase-shift \{$\phi_1=\phi_2\neq0, \pi$\} family. In {\bf (b)} the mixed 1 configuration is shown \{$\phi_1=\phi_2=0\ \text{or}\ 2\pi,\ \phi_3=\pi$\} or \{$\phi_1=\phi_2=\pi, \phi_3=0\ \text{or}\ 2\pi$\}. These families possess 2 positive and 1 negative $\chi_z$ each. In {\bf (c)} the mixed 2 family \{$\phi_1=\phi_3=\pi, \phi_2=0\ \text{or}\ 2\pi$\} is shown. It possesses 1 positive $\xz$ for $k<k_{cr}^{(2)}=1.0736$ and no positive $\xz$ for $k>k_{cr}^{(2)}$. At this point the mixed 2 family collides with the phase-shift one. In {\bf (d)} the phase-shift \{$\phi_1=\phi_2\neq0, \pi$\} family is shown and it is represented by $\phi_1=\phi_3=(3)$ and $\phi_2=(2)$ or $\phi_1=\phi_3=(4)$ and $\phi_2=(5)$. It exists for  $k_{cr}^{(1)}<k<k_{cr}^{(2)}$. At $k=k_{cr}^{(1)}$ it bifurcates from the anti-phase family, while at $k=k_{cr}^{(2)}$ it emerges with the mixed 2 family, and possesses no positive $\chi_z$. 
Fig.~\ref{fig:roots_4site_r2} contains all of the above families together, 
where, when two line segments coincide the more dense is shown.}
\label{pol_0_parts}
\end{figure}

By using the same arguments as in the previous section, which are also verified by our numerical investigation we have that for all the phase-shift breathers it is $\phi_1=\phi_3$. At $k=k_{cr}^{(1)}=0.3219$ the anti-phase  \{$\phi_1=\phi_2=\phi_3=\pi$\} family becomes subject to a bifurcation that generates the 
phase-shift 4-site breathers. We should also note
in passing (see details below) that, in addition to this supercritical
pitchfork, the figure reveals also a sub-critical pitchfork bifurcation that
terminates the two asymmetric branches upon their collision with the branch
with the mixed family \{$\phi_1=\phi_3=\pi$, $\phi_2=0$\} at $k_{cr}^{(2)} = 1.0736$.

\subsection{Stability of multibreathers}

The stability of the existing multibreather solutions can be calculated by using the previously developed theory. Their corresponding characteristic exponents are given to first order of approximation by (\ref{sz}). In the case under consideration of 4-site ($n=3$) breathers with range $r=2$ we have from (\ref{b1})
$$\hspace{-0.8cm}\frac{\pa^2\avh}{\pa\phi_i\pa\phi_j}=\left( \begin{array}{ccc}
f_1+k\,f_{1+2}&k\,f_{1+2}&0\\
k\,f_{1+2}&f_2+k\,f_{1+2}+k\,f_{2+3}&k\,f_{2+3}\\
0&k\,f_{2+3}&f_3+k\,f_{2+3}
\end{array}\right)
\quad\mathrm{and}\quad
{\bf L}=\left(\begin{array}{ccc}2&-1&0\\-1&2&-1\\0&-1&2\end{array}\right)$$
So, (\ref{z}) gives
$$
\ds{\bf Z}=\frac{\pa^2\avh}{\pa\phi_i\pa\phi_j}\cdot{\bf L}=
\left( \begin{array}{ccc}
2f_1+k f_{1+2}&kf_{1+2}-f_1&-kf_{1+2}\\
k(f_{1+2}-f_{2+3})-f_2&2f_2+k(f_{1+2}+f_{2+3})&k(f_{2+3}-f_{1+2})-f_2\\
-kf_{2+3}&kf_{2+3}-f_3&2f_3+kf_{2+3}
\end{array}\right)$$

For the general case  $\phi_1=\phi_3$, the eigenvalues of $\bf Z$ are
$${\chi_z}_1=2(f_1+kf_{1+2})\quad \text{and}\quad {\chi_z}_{2,3}=f_1+f_2+kf_{1+2}\pm\sqrt{f_1^2+f_2^2-2kf_1f_{1+2}+k^2f_{1+2}^2}.$$
The only configurations that are not included in the case above are the mixed ones \{$\phi_1=\phi_2=0,\ \phi_3=\pi$\} and
\{$\phi_1=\phi_2=\pi,\ \phi_3=0$\}, which both have 2 positive $\chi_z$, independently of the value of $k$.

\noindent {\bf Remark: } The ${\chi_z}_2$ eigenvalues of the anti-phase and mixed configurations can be used in order to calculate the values of $k_{cr}^{(1)}$ and $k_{cr}^{(2)}$. The $\xz_2$ for the anti-phase configuration is
$${\chi_z}_{2}=2f_\pi+kf_{0}+\sqrt{2f_\pi^2-2kf_\pi f_{0}+k^2f_{0}^2}.$$
Since for $k=k_{cr}^{(1)}$ it is ${\chi_z}_{2}=0$ we get $\ds k_{cr}^{(1)}=-\frac{f_\pi}{3f_0}\simeq\frac{1}{3}$. The last rough estimation can be performed only when we consider breathers with frequency $\w$ close to the phonon frequency $\w_p$ (see also the discussion in the previous section), where $|f_0|\simeq|f_\pi|$. In order to be more precise, for the potential and frequency used in the present work, we get $k_{cr}=0.3219$.

On the other hand, the ${\chi_z}_2$ eigenvalue for the mixed \{$\phi_1=\phi_3=\pi,\ \phi_2=0$\} configuration is
$${\chi_z}_2=f_0+(1+k)f_\pi+\sqrt{(1-k)^2f_\pi^2+f_0}.$$
Since for $k=k_{cr}^{(2)}$ it is ${\chi_z}_{2}=0$ we get $k_{cr}^{(2)}=\ds-\frac{f_0}{2f\pi+f_0}\simeq1$ or, for the specific potential and frequency used in this work $k_{cr}^{(2)}=1.0736$.\\

\noindent {\bf Remarks on the stability diagram of Figs. \ref{fig:roots_4site_r2} and \ref{pol_0_parts}:} In Figs.~\ref{fig:roots_4site_r2} and~\ref{pol_0_parts} all the multibreather families they exist in the present configuration ($n=3$, $r=2$) are shown. The kind of the line (or curve) used for every segment depends on the number of positive $\xz$ as follows: no positive $\xz\rightarrow$ solid, 1 positive $\xz\rightarrow$ dashed, 2 positive $\xz\rightarrow$ dashed-dotted, 3 positive $\xz\rightarrow$ dotted. All the families shown in Fig.\ref{pol_0_parts} are depicted together in Fig.\ref{fig:roots_4site_r2}. Since, if two segments coincide, the more dense is shown, in Fig.\ref{fig:roots_4site_r2} we can see only solid and dashed segments. The families which are depicted in these two figures are the following:

\begin{itemize}
\item \{$\phi_1=\phi_2=\phi_3=0\ (\text{or}\ 2\pi)$\} (in-phase). This family is shown in Fig.\ref{pol_0_parts}(a) and has 3 positive $\chi_z$.
\item \{$\phi_1=\phi_2=\phi_3=\pi$\} (anti-phase). It is shown in Fig.\ref{pol_0_parts}(a) and has no positive $\xz$ for $k<k_{cr}^{(1)}=0.3219$ while it possesses one positive $\xz$ for $k>k_{cr}^{(1)}$. At this point the anti-phase family bifurcates to provide the phase-shift breather family.
\item \{$\phi_1=\phi_2=0,\ \phi_3=\pi$\} or \{$\phi_1=\phi_2=\pi,\ \phi_3=0$\} (mixed 1). These families are depicted in Fig.\ref{pol_0_parts}(b) and they both possess 2 positive $\chi_z$
\item \{$\phi_1=\phi_3=\pi,\ \phi_2=0$ (or $2\pi$)\} (mixed 2). It is shown in Fig.\ref{pol_0_parts}(c). For $k<k_{cr}^{(2)}$ it has 1 positive $\chi_z$, while for $k>k_{cr}^{(2)}$ it has no positive $\xz$. At $k=k_{cr}^{(2)}=1.0736$, this family collides with the phase-shift family.
\item $\phi_1=\phi_3, \phi_2\neq0, \pi$ (phase shift). This family is represented in Fig.\ref{pol_0_parts}(d) by $\phi_1=\phi_3$=(3) and $\phi_2$=(2) (or $\phi_1=\phi_3$=(4) and $\phi_2$=(5)) and has no positive $\chi_z$. It begins to exist at $k=k_{cr}^{(1)}$ where it bifurcates from the anti-phase family and cease to exist at $k=k_{cr}^{(2)}$ where it collides with the mixed 1 family.
\end{itemize}

Since the stability of the multibreathers is also determined by the sign of $P\equiv\e\frac{\pa \w}{\pa J}$ the above are summarized, in terms of the linear stability of the corresponding configurations, in Table \ref{table2}.\\

\begin{table}[ht]
\begin{tabular}{|c|c|c|c|c|c|c|c}
\hline
 &$\hspace{0.5cm}P\hspace{0.5cm}$&$\hspace{0.7cm}k\hspace{0.7cm}$&\begin{tabular}{c}\\[-7pt] In-phase\\[-3pt] $\phi_1=\phi_2=\phi_3=0$\\[5pt] \end{tabular}&\begin{tabular}{c}Out-of-phase\\[-3pt] $\phi_1=\phi_2=\phi_3=\pi$\end{tabular}&\begin{tabular}{c} Phase-shift\\[-3pt] \mbox{$\phi_1=\phi_3, \phi_2\neq 0,\pi$}\end{tabular}&\begin{tabular}{c} Mixed\\[-3pt] $\phi_1=\phi_3=\pi, \phi_2=0$\end{tabular}\\
 \hline
\multirow{4}{*}{\begin{tabular}{c} \\[5pt]Linear\\ Stability\end{tabular}}&$P<0$&$k<k_{cr}^{(1)}$&unstable&stable&--&unstable\\
 & $P<0$ & $k_{cr}^{(1)}<k<k_{cr}^{(2)}$ & unstable & unstable & stable & unstable \\
 & $P<0$ & $k>k_{cr}^{(2)}$ & unstable & unstable & -- & stable \\
 & $P>0$ & $k<k_{cr}^{(1)}$ & stable & unstable & -- & unstable \\
 & $P>0$ & $k_{cr}^{(1)}<k<k_{cr}^{(2)}$ & stable & unstable & unstable & unstable \\
 & $P>0$ & $k>k_{cr}^{(2)}$ & stable & unstable & -- & unstable \\
 \hline
\end{tabular}
\caption{Stability of the various 4-site ($n=3$) and range $r=2$, breather configurations depending on the values of $P\equiv\frac{\pa \w}{\pa J}$ and $k$. With the dash we denote that this particular family does not exist for this range of values of $k$.}
\label{table2}
\end{table}

\section{4-site breathers with $r=3$}
The natural way to extend our study in 4-site breathers is to consider range of interaction $r=3$ (i.e., involving interactions with the 3 closest neighbors
on each side of the chain), 
in order for all the central oscillators to interact with each other.

\subsection{Persistence of multibreathers}
Bearing in mind that $\epsilon_i=\epsilon k_i$ and that
$k_1=1$, $\avh$ in this case becomes
$$\begin{array}{l}\ds\avh=-\frac{1}{2}\sum_{m=1}^\infty A_m^2\left\{ \cos(m\phi_1)+ \cos(m\phi_2)+\cos(m\phi_3)+\right.\\[8pt] \qquad\quad\left.+k_2\{\cos[m(\phi_1+\phi_2)]+\cos[m(\phi_2+\phi_3)]\}+ k_3\cos[m(\phi_1+\phi_2+\phi_3)] \right\},
\end{array}$$
while the corresponding persistence conditions become
$$\begin{array}{l}
M(\phi_1)+k_2M(\phi_1+\phi_2)+k_3M(\phi_1+\phi_2+\phi_3)=0\\[8pt]
M(\phi_2)+k_2\left[M(\phi_1+\phi_2)+M(\phi_2+\phi_3)\right]+k_3M(\phi_1+\phi_2+\phi_3)=0\\[8pt]
M(\phi_3)+k_2M(\phi_2+\phi_3)+k_3M(\phi_1+\phi_2+\phi_3)=0.
\end{array}$$

For every $k_i$, there exist the usual $\phi_i=0, \pi$ solutions, as well as
others as it can be seen in fig.~\ref{fig:roots_4site_r3_var_k2}. By keeping $k_3$ constant, we get various mono-parametric bifurcation diagrams
with $k_2$ as the parameter. Again, for all the phase shift configurations it is $\phi_1=\phi_3$. In fig.~\ref{fig:roots_4site_r3_var_k2}, the bifurcation diagrams for two values of $k_3$ are depicted, $k_3=0.2$ and $k_3=0.4$. We see that the value of ${k_2}_{cr}^{(1)}$ where the supercritical bifurcation occurs depends strongly on the value of $k_3$, while the value of ${k_2}_{cr}^{(2)}$ remains almost constant at ${k_2}_{cr}^{(2)}\simeq1.075$. The dependence of ${k_2}_{cr}$ with respect to $k_3$ is shown in fig. \ref{fig:k2cr_k3}.

Note that for $k_3\rightarrow 0$ this case coincides with the $r=2$ case
(i.e., the latter is a special case example)
and we retrieve the diagram of Fig.~\ref{fig:roots_4site_r2}.


\subsection{Stability}
As it has already mentioned the stability of the multibreathers is determined by the sign of the eigenvalues ${\chi_z}_i$, of matrix $\bf Z$ (\ref{z}). By (\ref{b1}) we get

$$\hspace{-1cm}
\ds\frac{\pa^2\avh}{\pa\phi_i\pa\phi_j}=
\left( \begin{array}{ccc}
f_1+k_2f_{1+2}+k_3f_{1+2+3}&
k_2f_{1+2}+k_3f_{1+2+3}&
k_3f_{1+2+3}\\
k_2f_{1+2}+k_3f_{1+2+3}&
f_2+k_2(f_{1+2}+f_{2+3})+k_3f_{1+2+3}&
k_2f_{2+3}+k_3f_{1+2+3}\\
k_3f_{1+2+3}&
k_2f_{2+3}+k_3f_{1+2+3}&
f_3+k_2f_{2+3}+k_3f_{1+2+3}
\end{array}\right),
$$
and since
$${\bf L}=\left(\begin{array}{ccc}2&-1&0\\-1&2&-1\\0&-1&2\end{array}\right)$$
we finally get
$$\hspace{-1.5cm}
\ds{\bf Z}=\frac{\pa^2\avh}{\pa\phi_i\pa\phi_j}\cdot{\bf L}=
\left( \begin{array}{ccc}
2f_1+k_2f_{1+2}+k_3f_{1+2+3}&
k_2f_{1+2}-f_1&
k_3f_{1+2+3}-k_2f_{1+2}\\
k_2f_{1+2}+k_3f_{1+2+3}-f_2-k_2f_{2+3}&
2f_2+k_2(f_{1+2}+f_{2+3})&
k_2f_{2+3}+k_3f_{1+2+3}-f_2-k_2f_{1+2}\\
k_3f_{1+2+3}-k_2f_{2+3}&
k_2f_{2+3}-f_3&
2f_3+k_2f_{2+3}+k_3f_{1+2+3}
\end{array}\right).
$$
Its eigenvalues are, for $\phi_1=\phi_3$ and non-specific values of $\phi_i$,
$${\chi_z}_1=2(f_1+k_2f_{1+2})$$
$$\hspace{-1.5cm}{\chi_z}_{2,3}=f_1+f_2+k_2f_{1+2}+k_3f_{2\phi_1+\phi_2}\pm\sqrt{f^2_1+f^2_2-2k_2f_1f_{1+2}+k_2^2f^2_{1+2}-2k_3f_2f_{2\phi_1+\phi_2}+k_3^2f^2_{2\phi_1+\phi_2}}.$$

\begin{figure}[htbp]
	\centering
		\includegraphics[width=8cm]{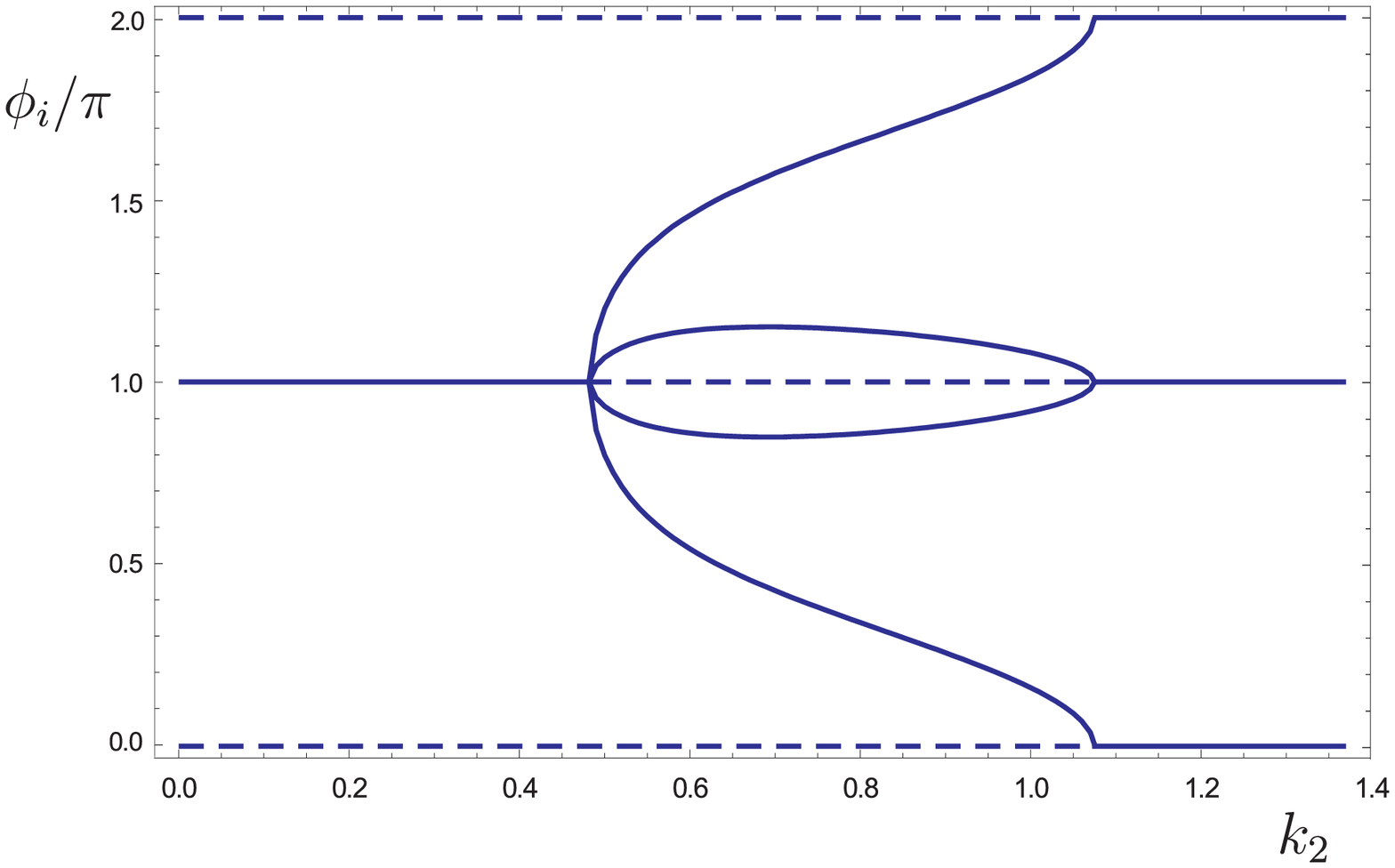}
		\hspace{0.5cm}
		\includegraphics[width=8cm]{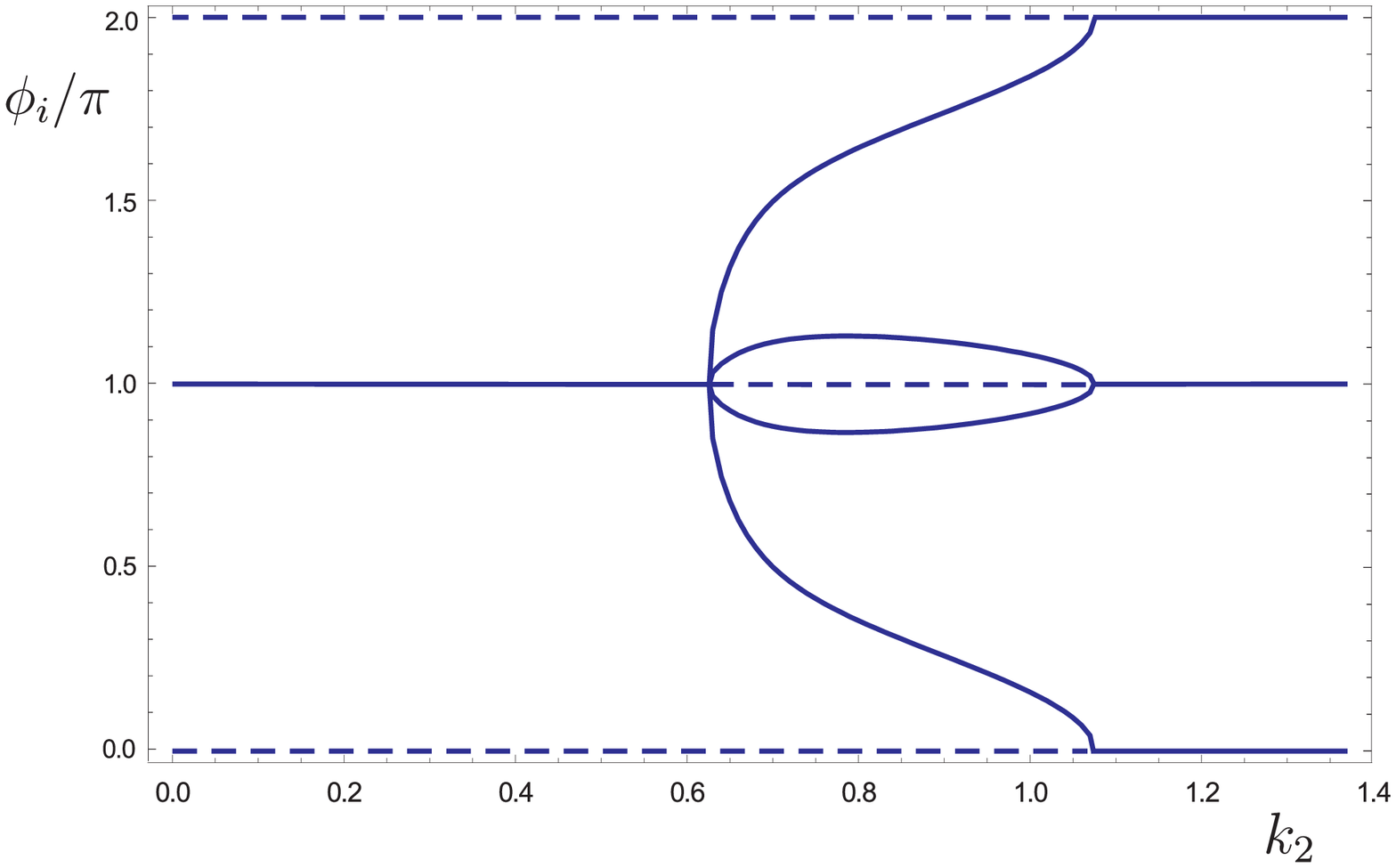}
	\caption{In these panels the various 4-site ($n=3$) multibreathers families in a chain with interaction range $r=3$ are shown. In the diagram we have considered $k_2$ variable, while $k_3=0.2$ in the left panel and $k_3=0.4$ in the right panel. The families depicted are qualitatively the same as the ones in the $n=3$, $r=2$ case. The only difference is that in the present case the value of ${k_2}_{cr}^{(1)}$ where the supercritical bifurcation occurs depends strongly on $k_3$ while the value of ${k_2}_{cr}^{(2)}$ where the subcritical bifurcation occurs remains almost fixed.}
	\label{fig:roots_4site_r3_var_k2}
\end{figure}

\begin{figure}[htbp]
	\centering
		\includegraphics[width=9cm]{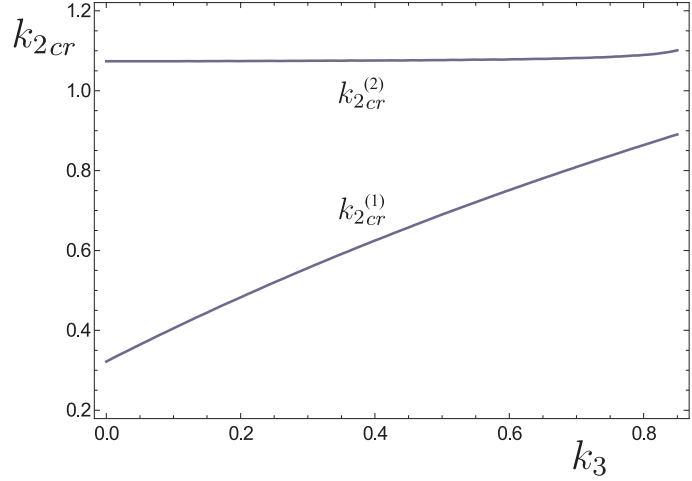}
	\caption{The values of the two ${k_2}_{cr}$, where the bifurcations occur, with respect to $k_3$. Although ${k_2}_{cr}^{(1)}$ depends strongly on $k_3$, ${k_2}_{cr}^{(2)}$ remains almost constant at ${k_2}_{cr}^{(2)}\simeq1.075$.}
	\label{fig:k2cr_k3}
\end{figure}

Although such analytical formulas exist and accurately predict the stability
and bifurcations of the system, a clearer understanding emerges from the
observation of the associated bifurcation diagrams (fig.~\ref{fig:roots_4site_r3_var_k2}). The diagrams present exactly the same solution families as in Fig.~\ref{fig:roots_4site_r2}, but in this case the value of ${k_2}_{cr}^{(1)}$, where the supercritical bifurcation occurs, is strongly affected by the value of $k_3$, while, the value of ${k_2}_{cr}^{(2)}$, where the subcritical
bifurcation occur,s remains almost constant with ${k_2}_{cr}^{(2)}\simeq1.075$ (Fig.~\ref{fig:k2cr_k3}). This indicates that,
for the range of values of $k_3$ considered in this figure, the parametric interval
of $k_2$ (the strength of next-nearest neighbor interactions) 
over which phase-shift solutions exist narrows as $k_3$ (the strength
of interaction with the third-nearest-neighbors) is increased.

\section{4-site breathers in a 2D square lattice with $r=2$}
\begin{figure}[htbp]
	\centering
		\includegraphics[width=8cm]{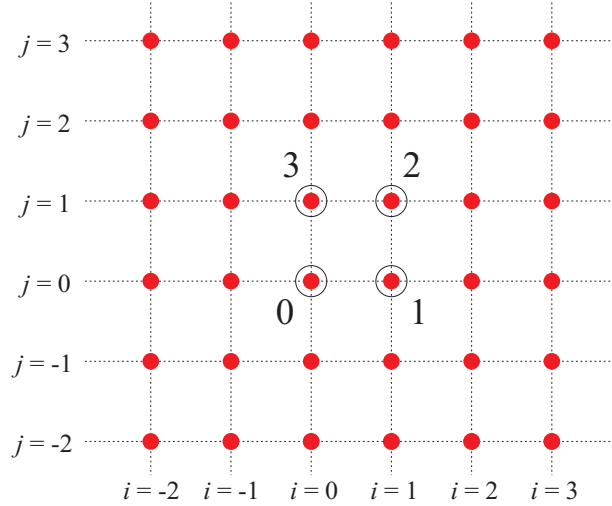}
	\caption{In the square lattice under consideration each oscillator is coupled with its neighbour not only in the horizontal and vertical directions but in the diagonal directions as well. So, every lattice site interacts with its 8 neighbours instead of the 4 of the classical KG square lattice configuration. In this setting, we consider 4-site breathers. Let the encircled oscillators in the figure be the central oscillators which are denoted as 0, 1, 2, 3 respectively}
	\label{fig:square_lattice}
\end{figure}

We now turn our considerations to the case of a square lattice, as the one in
Fig.~\ref{fig:square_lattice}, with nearest-neighbor interactions, not only with the horizontal and vertical neighbors, but with the diagonal as well.
The latter interaction is assumed
to have a strength $\epsilon_2=k \epsilon_1$ (where
$\epsilon_1 \equiv \epsilon$ will be taken to denote the coupling
strength of adjacent nodes along the lattice axes).
The Hamiltonian for this system is
\beq \begin{array}{cll}H&=\ds\sum_{i=-\infty}^{\infty}\sum_{j=-\infty}^{\infty}\left[ \frac{1}{2}p_{ij}^2 +V(x_{ij})\right]&+\ds\frac{\e_1}{2}\sum_{i=-\infty}^{\infty}\sum_{j=-\infty}^{\infty}\left[(x_{ij}-x_{i-1,j})^2+(x_{ij}-x_{i,j-1})^2\right]\\
& &+\ds\frac{\e_2}{2}\sum_{i=-\infty}^{\infty}\sum_{j=-\infty}^{\infty}\left[(x_{ij}-x_{i-1,j-1})^2+(x_{ij}-x_{i-1,j+1})^2\right]
\end{array}
\label{sq_r2}\eeq

or

\beq \begin{array}{cl}H=H_0+\e H_1&=\ds\sum_{i=-\infty}^{\infty}\sum_{j=-\infty}^{\infty}\left[ \frac{1}{2}p_{ij}^2 +V(x_{ij})\right]+\\
&+\ds\frac{\e}{2}\sum_{i=-\infty}^{\infty}\sum_{j=-\infty}^{\infty}\left\{(x_{ij}-x_{i-1,j})^2+(x_{ij}-x_{i,j-1})^2+\ds  k\left[(x_{ij}-x_{i-1,j-1})^2+(x_{ij}-x_{i-1,j+1})^2\right]\right\}
\end{array}
\label{sq_r2_k}\eeq
We consider 4 ``central'' oscillators in the anti-continuum limit and we denote them by 0, 1,2 ,3 as it is shown in fig.\ref{fig:square_lattice}. We have then $\phi_1=w_1-w_0$, $\phi_2=w_2-w_1$, $\phi_3=w_3-w_2$ and $\phi_4=w_0-w_3$. But since, by construction, we have $\phi_4=2\pi-\phi_1-\phi_2-\phi_3$, we have finally only 3 independent $\phi_i$'s. The $\avh$ in this case is
$$\begin{array}{l}\ds\avh=-\frac{1}{2}\sum_{m=1}^\infty A_m^2\left\{ \cos(m\phi_1)+ \cos(m\phi_2)+\cos(m\phi_3)+\cos[m(\phi_1+\phi_2+\phi_3)]\right.\\[8pt] \qquad\qquad\qquad\qquad\quad\left.+k\{\cos[m(\phi_1+\phi_2)]+\cos[m(\phi_2+\phi_3)]\} \right\}\end{array}$$
and the corresponding persistence conditions are
$$\begin{array}{ccl}
M(\phi_1)+M(\phi_1+\phi_2+\phi_3)+k\,M(\phi_1+\phi_2)&=&0\\[8pt]
M(\phi_2)+M(\phi_1+\phi_2+\phi_3)+k[M(\phi_1+\phi_2)+M(\phi_2+\phi_3)]&=&0\\[8pt]
M(\phi_3)+M(\phi_1+\phi_2+\phi_3)]+k\,M(\phi_2+\phi_3)&=&0
\end{array}$$

\begin{figure}[htbp]
	\centering
		\includegraphics[width=12cm]{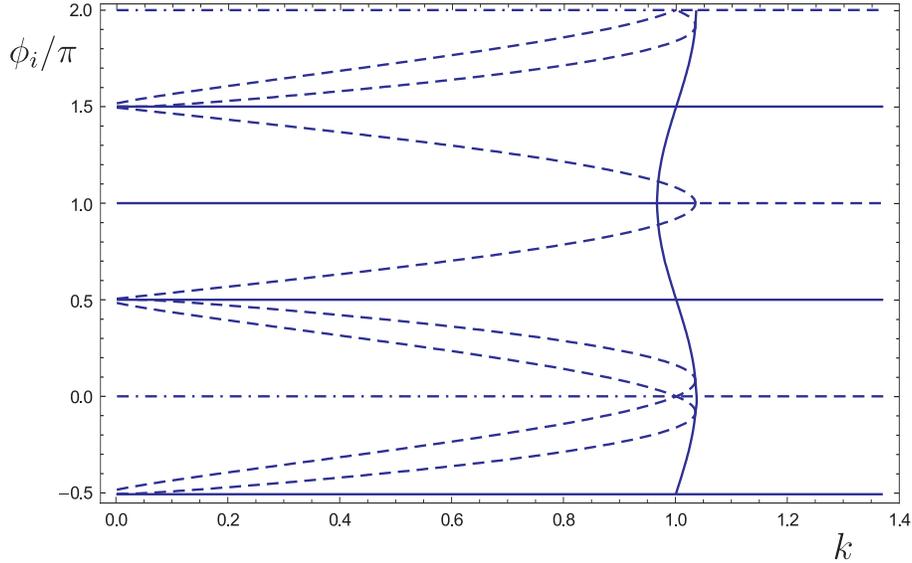}
	\caption{In this diagram, the various multibreather families of the square lattice configuration, are depicted. The value of $k$ determines the strength of the coupling in the diagonal direction. For a more detailed description of the particular families one can also refer to Fig. \ref{pol_1_parts}.}
	\label{fig:roots_square}
\end{figure}

This case coincides with the 1D 4-site $r=3$ chain case, with $k_3=k_1=1$ and $k_2=k$. Hence, the results (both for the persistence and the stability of the
solutions) are a special case of the previous section. All the existing multibreather families of this configuration are depicted in fig.~\ref{fig:roots_square}.

In this diagram we can observe the appearance of \{$\phi_1=\phi_2=\phi_3=\pi/2$\} family, which is the
vortex solution of the classical square Klein-Gordon lattice; see
e.g., the relevant discussion in~\cite{IJBC}. In addition, there are several phase-shift families, the stability of which will be analyzed below. Interestingly, all these phase-shift breathers cease to exist at a critical value of $k=k_{cr}^{(2)}=1.03549$ except of the vortex one.

\begin{figure}[htb]
\begin{tabular}{ccccc}
\includegraphics[width=8cm]{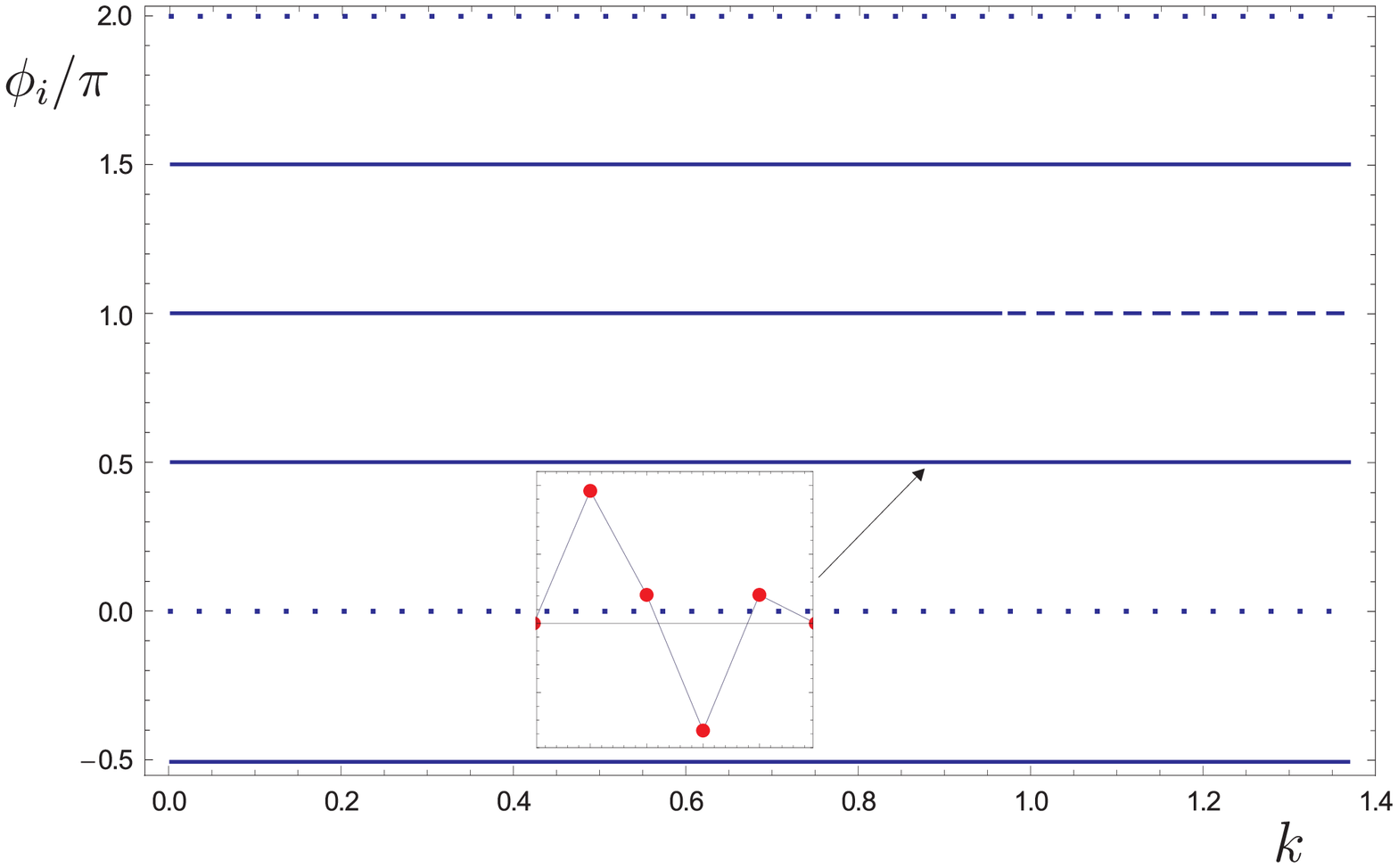}&$\hspace{0.5cm}$&\includegraphics[width=8cm]{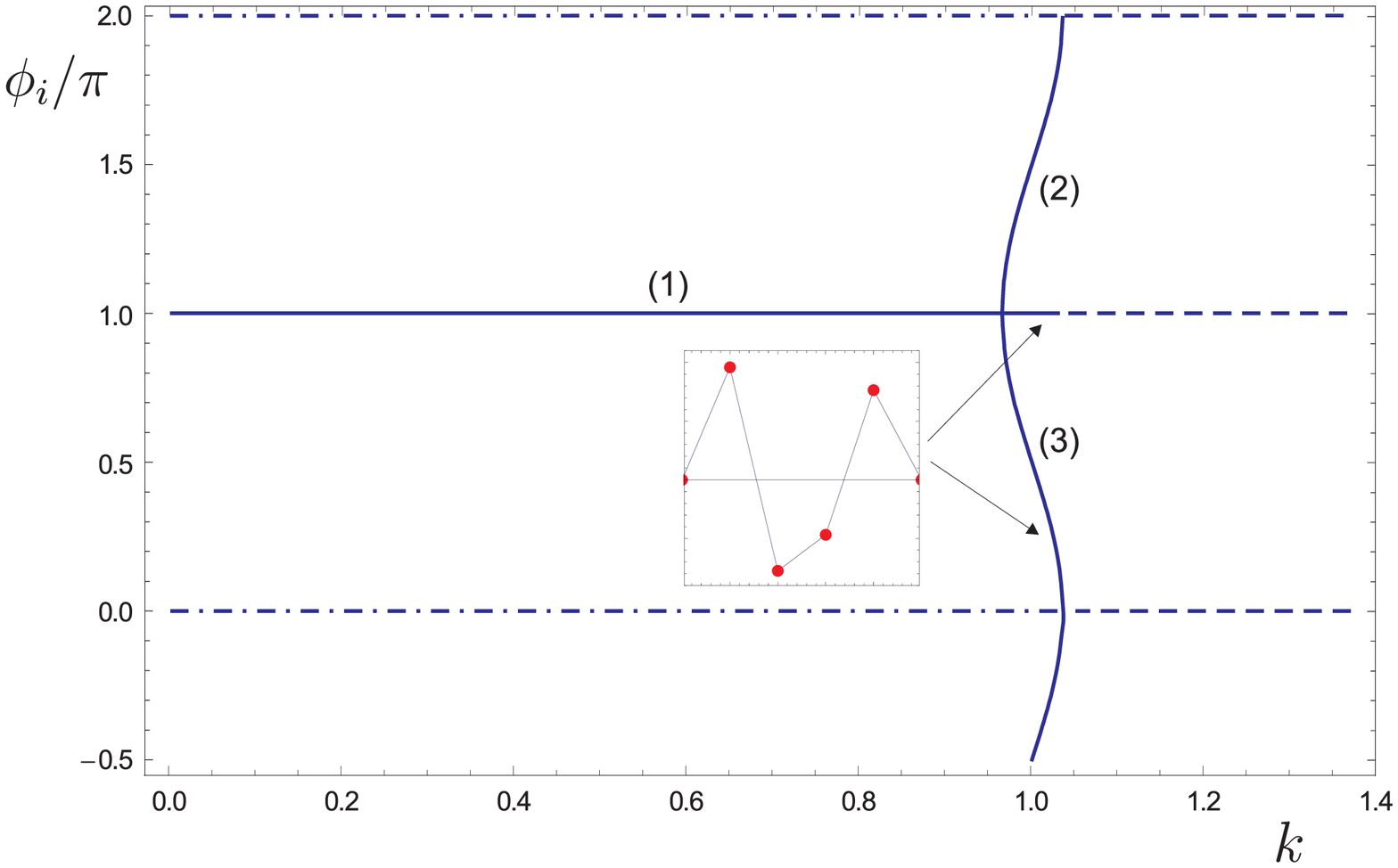}\\[-5pt]
(a)& &(b)\\[10pt]
\includegraphics[width=8cm]{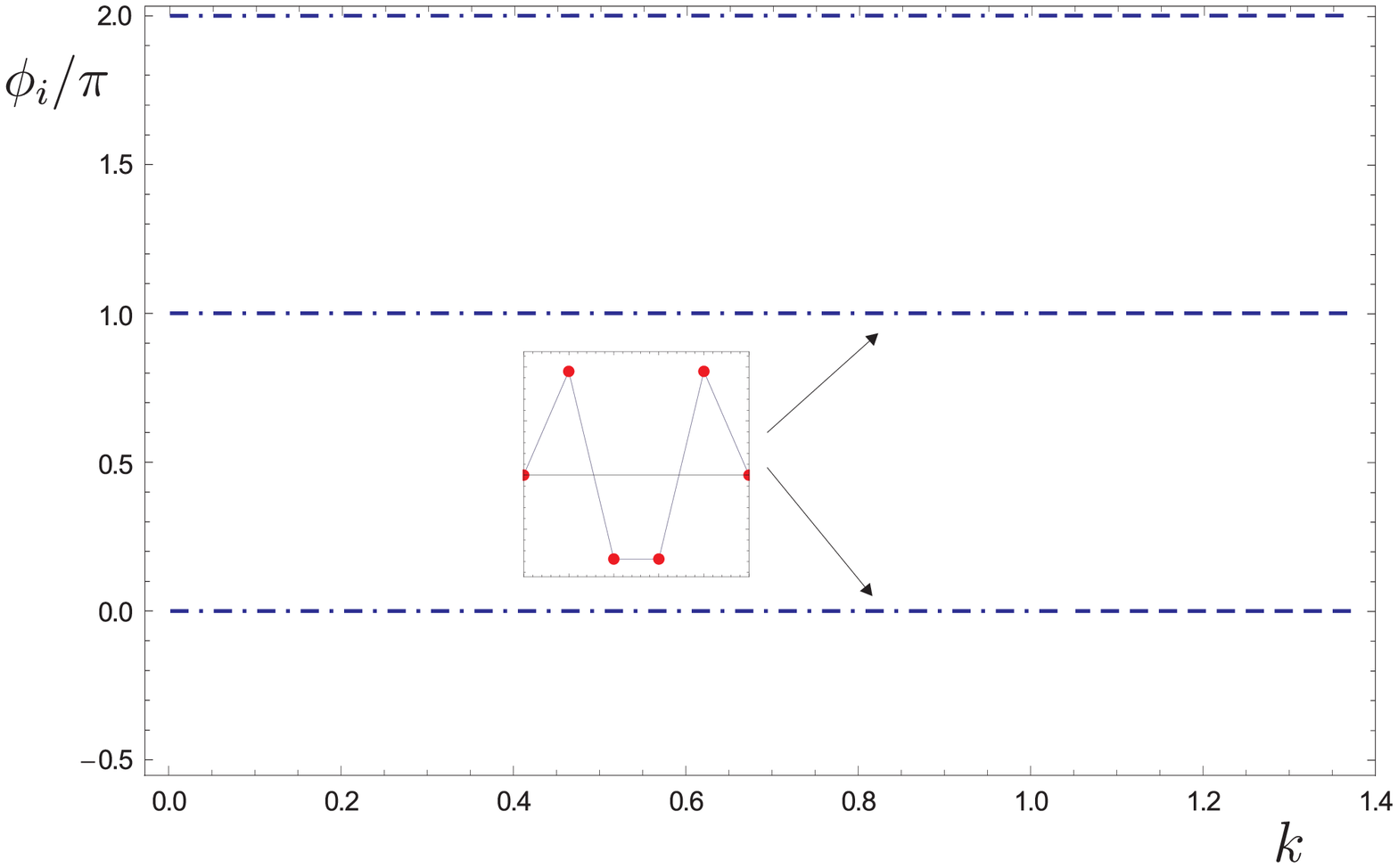}&$\hspace{0.5cm}$&\includegraphics[width=8cm]{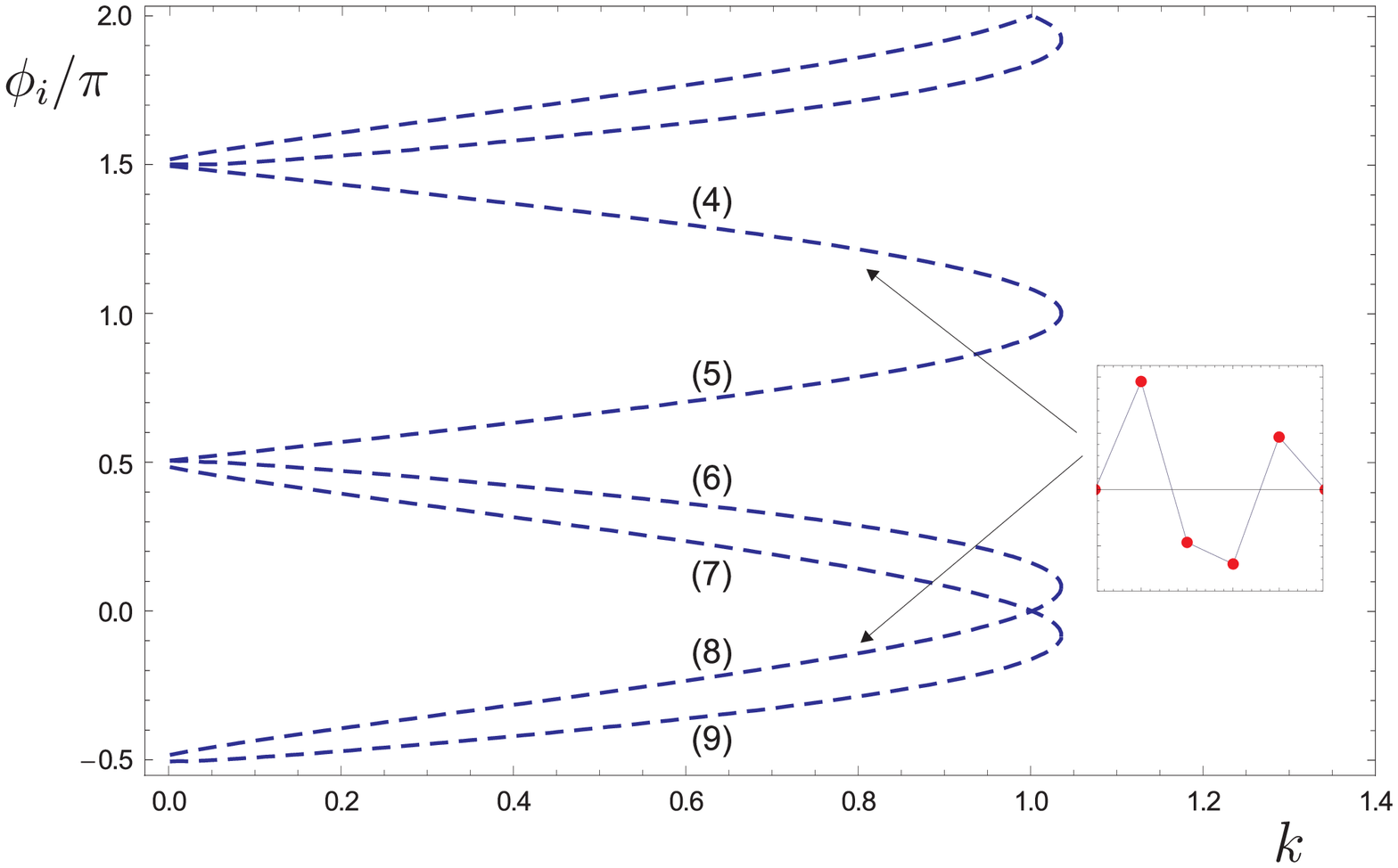}\\[-5pt]
(c)& &(d)
\end{tabular}
\caption{The various multibreather families that consist Fig. \ref{fig:roots_square} are depicted. Portraits of the configuration of the central oscillators 
for the different multibreather families are shown as insets in the 
corresponding diagrams. Note that, for better presentation, instead of showing the the 4 central oscillators in a square configuration, we show their 1D equivalent. The line (or curve) type in the figures depends on the number of positive $\xz$ the corresponding family possesses: no positive $\xz$ corresponds to solid line, 1 positive $\xz$ corresponds to dashed line, 2 to dashed-dotted and 3 to dotted line. In {\bf (a)} three families are shown. The first family is the in-phase one \{$\phi_1=\phi_2=\phi_3=0\ \text{or}\ 2\pi$\}, which possesses 3 positive $\xz$. The second family is the vortex \{$\phi_1=\phi_2=\phi_3=\pi/2\ (\text{or}\ 3\pi/2)$\} family. It possesses no positive $\xz$. The third family is the anti-phase one \{$\phi_1=\phi_2=\phi_3=\pi$\}. This family possesses no positive $\xz$ for $k<k_{cr}^{(1)}=0.96572$, while for $k>k_{cr}^{(1)}$ it acquires 2 positive $\xz$. At this point it bifurcates to provide the phase-shift 1 family. 
In this figure only the inset of the vortex family is present because all the others are the same as the ones presented for the equivalent 1D configuration, in figs. \ref{pol_0_parts}. In {\bf (b)} the phase-shift 1 \{$\phi_1=\phi_3=\pi$, $\phi_2=\phi$\} family is shown, which exists for $k_{cr}^{(1)}<k<k_{cr}^{(3)}=1.03549$. It is depicted by $\phi_1=\phi_3=(1)$ and $\phi_2=(2)$ or $\phi_1=\phi_3=(1)$ and $\phi_2=(3)$.  It possesses no positive $\xz$ for $k_{cr}^{(1)}<k<k_{cr}^{(2)}=1.0344$. At $k=k_{cr}^{(2)}$ it collides with the phase-shift 2 family. For $k_{cr}^{(2)}<k<k_{cr}^{(3)}$ the family possesses 1 positive $\xz$. At $k=k_{cr}^{(3)}$ it collides with the mixed family. In {\bf (c)} the mixed family \{$\phi_1=\phi_3=\pi$, $\phi_2=0$\} is depicted. It has 2 positive $\xz$ for $k<k_{cr}^{(2)}$ and 1 positive $\xz$ for $k_2>k_{cr}^{(3)}$. Finally in {\bf (d)} two phase-shift families are depicted. The phase phase-shift 2a family, which is represented by $\phi_1=\phi_3=(4)$ and $\phi_2=$(8) which collides with the $\phi_1=\phi_3=(5)$ and $\phi_2=$(6) for $k_2=k_{cr}^{(2)}$, and the phase-shift 2b family which is represented by $\phi_1=\phi_3=(4)$ and $\phi_2=$(9) which collides with the $\phi_1=\phi_3=(5)$ and $\phi_2=$(7) for $k_2=k_{cr}^{(2)}$. They both possess 1 positive $\xz$.
}
\label{pol_1_parts}
\end{figure}

The mutibreather families that are supported by the present configuration are described in what follows. 

\begin{itemize}
\item \{$\phi_1=\phi_2=\phi_3=0$\} (in-phase) It is shown in Fig.\ref{pol_1_parts}(a). It possesses 3 positive $\xz$ independently of the value of $k_2$.
\item \{$\phi_1=\phi_2=\phi_3=\pi/2$\} (vortex). It is shown in Fig.\ref{pol_1_parts}(a). It has no positive $\xz$ independently of the value of $k_2$ and does not interact (i.e., collide) with any other family of solutions.
\item \{$\phi_1=\phi_2=\phi_3=\pi$\} (anti-phase). It is shown in Fig.\ref{pol_1_parts}(a). It has no positive $\xz$ for $k<k_{cr}^{(1)}=0.96572$, while for $k>k_{cr}^{(1)}$ it acquires 2 positive $\xz$.
\item \{$\phi_1=\phi_3=\pi$, $\phi_2=\phi$\} (phase-shift 1). It is represented by in Fig.\ref{pol_1_parts}(b) $\phi_1=\phi_3=(1)$ and $\phi_2=(2)$ or $\phi_1=\phi_3=(1)$ and $\phi_2=(3)$. It exists for $k_{cr}^{(1)}<k_2<k_{cr}^{(3)}$. It bifurcates from the anti-phase family at $k=k_{cr}^{(1)}$ and possesses 3 negative $\xz$ for $k_{cr}^{(1)}<k<k_{cr}^{(2)}=1.0344$. At $k_2=k_{cr}^{(2)}$ it collides with the phase-shift 2 family in a bubcritical pitchfork bifurcation. So, for $k_{cr}^{(2)}<k<k_{cr}^{(3)}$ it possesses 2 negative and a positive $\xz$. At $k=k_{cr}^{(3)}$ it collides with the mixed family.
\item \{$\phi_1=\phi_3=\pi$, $\phi_2=0$\} (mixed). It is shown in Fig.\ref{pol_1_parts}(c). It has 2 positive $\xz$ for $k<k_{cr}^{(3)}=1.0355$ and 1 positive $\xz$ for $k_2>k_{cr}^{(3)}$.
\item (phase-shift 2a) It is represented in Fig.\ref{pol_1_parts}(d) by $\phi_1=\phi_3=(4)$ and $\phi_2=$(8) which collides with the $\phi_1=\phi_3=(5)$ and $\phi_2=$(6) for $k_2=k_{cr}^{(2)}$ and have 1 positive $\xz$.
\item (phase-shift 2b) It is represented in Fig.\ref{pol_1_parts}(d) by $\phi_1=\phi_3=(4)$ and $\phi_2=$(9) which collides with the $\phi_1=\phi_3=(5)$ and $\phi_2=$(7) for $k_2=k_{cr}^{(2)}$ and have 1 positive $\xz$.
\end{itemize}
\noindent{\bf Notes on Figs. \ref{fig:roots_square} and \ref{pol_1_parts}:} 
\begin{enumerate}
\item At $k=k_{cr}^{(2)}$ the $\phi_2=(3)$ branch of the phase-shift 1 family collides with the phase-shift 2a family, while the the $\phi_2=(2)$ branch of the phase-shift 1 family collides with the symmetric (and equivalent) of branches $\phi_2=(7)$ and $\phi_2=(9)$ of the phase-shift 2b family.
\item  At $k=0$ the phase-shift 2 family approaches very much the vortex family, so it is plausible to expect that at that point they coincide. Yet, there is a small difference in the values of $\phi_i$ of the two families due to the nonlinear character of the oscillators constituting the lattice, i.e., due to the existence of an infinity of terms in the development (\ref{xdevel}). For a smaller frequency of oscillation, the nonlinear character of the oscillation becomes stronger, hence the terms after the first in (\ref{xdevel})  become larger and the two families are more clearly separated.
\end{enumerate}
\section{Discussion - Comparison with the DNLS results}
It should be noted here that a number of results similar to the ones
presented herein have been recently presented in the context
of the DNLS equation e.g. in~\cite{pgkpla,chong}. The setting of
the DNLS essentially reflects a special case example of our
Klein-Gordon calculation where instead of the existence and
stability conditions reflecting a sum over {\it all} the harmonics
due to the U$(1)$ invariance of the underlying model, only the
first harmonic is present. Nevertheless, the latter is sufficient
to induce a number of the conclusions that we inferred herein.
In particular, next-nearest neighbor 
interactions create phase-shift multibreathers
(which were also parallelized to discrete vortex breathers in
hexagonal lattices), as illustrated in~\cite{pgkpla}. As also
shown in the same work, the long range interactions may drastically affect
the stability properties of two-dimensional discrete vortices
(in square lattices). On the other hand, the work of~\cite{chong}
provided a different analytical handle, via variational
approximations, on the solutions that arise in settings
with long range interactions. Furthermore, it was able to capture phenomena
(both analytically and numerically)
such as the supercritical or subcritical pitchfork bifurcations
for such phase-shift multibreather solutions with NNN interactions.
For instance, in the
DNLS case the supecritical bifurcation leading to the emergence
of such solutions would happen precisely at $k=0.5$ (due to the
relevance of just the first harmonic) and not at $k=0.48286$
as obtained here in section IV B for the Klein-Gordon case. Nevertheless, the
basic phenomenology remains intact.

\section{Conclusions}
Classical Klein-Gordon chain with nearest neighbor
interactions support multibreather solutions only with phase differences
between successive oscillators of $\phi_i=0, \pi$. There, the stability
scenaria are specific and well known. For a KG chain with
$P=\e\frac{\pa \w}{\pa J}<0$ the anti-phase configuration is the only stable
one, while for $P=\e\frac{\pa \w}{\pa J}>0$ the in-phase configuration is
the only stable multibreather solution.

On the other hand, in chains with long range interactions the picture is
{\it substantially} different. First of all, in such chains, multibreathers
with $\phi_i\neq0, \pi$ (phase-shift multibreathers) can be supported in
addition to the standard $\phi_i=0, \pi$ ones. The existence of phase-shift
multibreathers as well as the specific $\phi_i$'s of such profiles
depend on the various coupling parameters $\e_i$ within the chain.
There are critical values of $k_i=\e_i/\e_1$ past which a bifurcation occurs
(typically a supercritical pitchfork)
and phase shift breathers begin to exist. Past this bifurcation 
point, the stability
properties of the existing multibreathers are significantly modified,
although this also depends on the particular (soft or hard) nature of
the nonlinearity. As, however, additional parameters are tuned (e.g.,
higher ranges of neighbor interactions), it is also possible for
such phase-shift solutions to terminate in subcritical pitchfork bifurcations.

These results are not unique to the realm of one-dimensional lattices
with higher range of interactions. They can also be developed for
two-dimensional square lattices in which case they may lead to bifurcations
or terminations of the families of discrete vortices which arise therein.
Such vortices sustained by the two-dimensional analogs of the lattice
can be of either a symmetric or asymmetric type. In particular, in the
case considered herein, the presence of diagonal coupling within the
square was critical in inducing the emergence of asymmetric such patterns.

This study opens a number of a directions for further investigation.
Firstly, it would be very relevant to examine particular
functional forms of the decay of the long range interactions
(e.g., exponentially or polynomially decaying ones) to identify
whether any systematic conclusions can be derived on the basis of such
decay laws. Secondly, it would also be very interesting
to examine the interplay of the geometry of higher dimensional
lattices (and the interactions that they present) with the strength
of the long range interactions that can be considered therein and
to try to derive some general conclusions
about the possible stable/unstable discrete soliton and discrete vortex
solutions. Finally, an important and immediate direction that
can be followed with the results of the paper in hand could be the effect 
of long-range interaction in phase-shift phonobreathers, whose stability 
for nearest-neighbor interaction was considered in \cite{CAR11}. A physical 
application of relevance and worthwhile of further investigation concerns 
the biological models for DNA \cite{CAGR02,CSAH04} or protein alpha-helices \cite{AGCC02}, where dipole long-range interactions are relevant.

\acknowledgments

The contribution of VK and VR in this research has been partially co-financed by the European
Union (European Social Fund – ESF) and Greek national funds through the Operational
Program "Education and Lifelong Learning" of the National Strategic Reference Framework
(NSRF) - Research Funding Program: THALES. Investing in knowledge society through the
European Social Fund.

PGK gratefully acknowledges
support from the National Science Foundation, under grants
DMS-0806762, CMMI-1000337, and from the Alexander von Humboldt
Foundation as well as from the Alexander S. Onassis Public Benefit Foundation.

JC acknowledges financial support from the MICINN project FIS2008-04848.

\bibliographystyle{plain}
\bibliography{lri}

\end{document}